\documentclass[aps,pra,showpacs,onecolumn, superscriptaddress, longbibliography]{revtex4-2}
\usepackage{lineno}
\usepackage{xcolor}
\usepackage[normalem]{ulem} 
\usepackage{graphicx}
\usepackage{dcolumn}
\usepackage{bm}
\usepackage[colorlinks]{hyperref}
\usepackage{amsmath,braket}
\usepackage{mathtools}

\makeatletter
\renewcommand\@makefnmark{\hbox{\@textsuperscript{\normalfont\@thefnmark}}}
\makeatother
\usepackage{orcidlink}

\begin{document}
\title{Crossovers between Time-Bin and Energy-Time Entanglement in Homodyned Resonance Fluorescence}

\newcommand{\orcidauthorA}{0000-0000-0000-000X} 
\newcommand{\orcidauthorC}{0000-0003-2251-9316} 
\newcommand{\orcidauthorD}{0000-0001-5276-9151} 

\author{Zi-Qi Zeng}
\email{Corresponding author:zengzq@baqis.ac.cn}
\affiliation{Beijing Academy of Quantum Information Sciences, Beijing 100193, China}
\affiliation{Beijing National Laboratory for Condensed Matter Physics, Institute of Physics, Chinese Academy of Sciences, Beijing 100190, China}
\affiliation{University of Chinese Academy of Sciences, Beijing 100049, China}

\author{Jian Wang}
\affiliation{Beijing Academy of Quantum Information Sciences, Beijing 100193, China}

\author{Bang Wu\orcidlink{0000-0003-2251-9316}}

\affiliation{Beijing Academy of Quantum Information Sciences, Beijing 100193, China}

\author{Zhiliang Yuan\orcidlink{0000-0001-5276-9151}}
\affiliation{Beijing Academy of Quantum Information Sciences, Beijing 100193, China}

\date{\today}
\begin{abstract}
We theoretically investigate photonic entanglement generated by homodyning resonance fluorescence from a two-level emitter with a phase-locked laser field. Under weak excitation, increasing the relative laser-field amplitude produces crossovers from time-bin to energy--time and back to time-bin entanglement. The intermediate energy--time regime occurs near amplitude matching, where destructive interference suppresses the single-photon component, allowing the two-photon contributions to dominate the coincidence signal. We evaluate the resulting photon statistics and first-order coherence and characterize the entanglement through violations of the CHSH Bell inequality in a Franson interferometer. This scheme provides a route to controlling photonic entanglement through linear optics and coherent interference.
\end{abstract}

\maketitle

\section{Introduction}

Quantum entanglement is a fundamental resource in quantum information science
\cite{Bennett1993Teleportation,Bennett1996Purification}, underpinning applications in quantum communication
\cite{Ekert1991PRL,Brendel1999PRL,Duan2001Nature}, quantum computation
\cite{Bennett2000,Knill2001Nature,Kok2007RMP}, and quantum metrology
\cite{Giovannetti2011NP
}. Photonic entanglement is conventionally generated through spontaneous parametric down-conversion and other nonlinear optical processes
\cite{Kwiat1995PRL,Li2005,Takesue2005}, which have provided a highly successful platform for quantum optics and photonic quantum information processing
\cite{Bouwmeester1997Nature,Walther2005Nature,Pan2012RMP,Zhong2018PRL}. 

A conceptually different route is offered by resonance fluorescence (RF) from a coherently driven two-level emitter. RF-based entanglement has been explored through several approaches, including single-photon path entanglement~\cite{Zeng2025Bell}, spectral filtering~\cite{Peiris2017PRL,Liu2024NP}, and interferometric schemes~\cite{AMZI2026}.
In the weak-excitation (Heitler) regime, RF exhibits strong photon antibunching on the timescale of the radiative lifetime while retaining phase coherence with the driving laser and inheriting its narrow linewidth
\cite{hoffges1997heterodyne,nguyen2011ultra,mattiesen2012,LopezCarreno2018,Wang2025NC,Huang2026PRA}. In the description adopted here, this coexistence of single-photon character and optical phase coherence provides a natural basis for engineering quantum states through interference.

Recently, self-homodyne interference 
of RF in an asymmetric Mach--Zehnder interferometer (AMZI) has been demonstrated as an all-linear-optical route to entanglement synthesis~\cite{AMZI2026}. Destructive interference suppresses the single-photon component, allowing the surviving two-photon components to exhibit high-visibility Franson interference~\cite{Franson1989} and violate the Clauser--Horne--Shimony--Holt (CHSH) Bell inequality~\cite{Clauser1969PRL}. Interference of RF with a coherent laser field was originally proposed as a means of tailoring its photon-number statistics~\cite{LopezCarreno2018}, and has since been used to manipulate photon correlations~\cite{Kim2025SA,Bracht2026OpticaQ,Yin2026OL}. Whether such RF--laser interference can also be used to generate and continuously transform photonic entanglement remains less well explored.

In this work, we theoretically investigate the entanglement generated by
homodyning RF with a phase-locked coherent field and analyzing the resulting field with a Franson interferometer. We derive the coincidence correlations and the corresponding CHSH Bell parameter $S$. As the coherent-field amplitude is varied relative to the RF field, the system undergoes a continuous crossover from time-bin entanglement to energy--time entanglement and back to time-bin entanglement, before approaching classical coherent-field interference at high coherent-field strength. The crossover arises from photon-number-selective destructive interference, whereby different Fock-state components are suppressed at different amplitude ratios. These results establish homodyned RF as a tunable route to entanglement synthesis and suggest a pathway toward controlling higher-order multiphoton states through the same interference mechanism.

\section{Theoretical framework}

\subsection{Pure-state description of resonance fluorescence}

For a coherently driven two-level emitter under steady-state conditions,
in the absence of pure dephasing and at exact resonance, the joint
emitter--field state at a given time can be written as
\cite{Wang2025NC}
\begin{equation}
|\psi\rangle_{ap}
=
\sqrt{p_0}|0\rangle|g\rangle
+
\sqrt{p_1}e^{-i\omega_L t}
\frac{|0\rangle|e\rangle+|1\rangle|g\rangle}{\sqrt{2}} .
\label{eq:joint_state}
\end{equation}
Here, the subscripts $a$ and $p$ refer to the emitter and photonic-field
subsystems, respectively. The states $|g\rangle$ and $|e\rangle$ denote
the ground and excited states of the emitter, while $|0\rangle$ and
$|1\rangle$ denote the vacuum and single-photon states of the fluorescence
field.  $\omega_L$ is the laser angular frequency. The probabilities $p_0$ and $p_1$ are given by
\begin{align}
p_0 &= \frac{1}{1+2\Omega^2T_1^2}, &
p_1 &= \frac{2\Omega^2T_1^2}{1+2\Omega^2T_1^2},
\end{align}
with $p_0+p_1=1$, where $\Omega$ is the Rabi frequency and $T_1$ is the radiative lifetime. 
The dimensionless excitation strength $\Omega^2T_1^2$
is related to the incident laser power through
$\eta_{\rm ab}\bar{n}=\Omega^2T_1^2$, where $\bar{n}$ is the mean number
of incident photons within one radiative lifetime $T_1$, and
$\eta_{\rm ab}$ denotes the emitter absorption efficiency in the weak-drive limit. Consequently,
the parameter $p_1$ increases with excitation power, from $p_1\rightarrow 0$
in the Heitler regime to $p_1\rightarrow 1$ in the strong-driving limit.

The joint state provides an intuitive picture of how RF exhibits both antibunching and first-order coherence. Its photonic subsystem is restricted to the vacuum and single-photon Fock states, ensuring perfect antibunching, $g^{(2)}(0)=0$. Coherence between $|0\rangle|g\rangle$ and $(|0\rangle|e\rangle+|1\rangle|g\rangle)/\sqrt{2}$ establishes the phase coherence of the fluorescence field, with
\begin{equation}
|g^{(1)}(\tau)|=p_0
\label{eq:g1}
\end{equation}
for delays much longer than $T_1$ but within the coherence time of the driving laser. Thus, in the Heitler regime ($p_0\to1$), the long-delay first-order coherence approaches unity, corresponding to a laser-like linewidth. As the excitation power increases, $p_0$ decreases and the long-delay coherence gradually diminishes.

Tracing over the emitter from the joint state yields a mixed reduced state of the fluorescence field:
$\rho_p=\frac{2p_0+p_1}{2}|0\rangle\langle 0|+\frac{p_1}{2}|1\rangle\langle 1|+\sqrt{\frac{p_0p_1}{2}}\left(|0\rangle\langle 1|+|1\rangle\langle 0|\right)$.
 This state is exactly reproduced by passing the pure state
\begin{equation}
|\varphi\rangle
=
\sqrt{p_0}|0\rangle
+
\sqrt{p_1}e^{-i\omega_L t}|1\rangle
\label{eq:pure_RF}
\end{equation}
through a $3$\,dB loss channel and tracing over the loss mode~\cite{Wang2025NC}. Since normalized correlation functions are invariant under linear loss, Eq.~\eqref{eq:pure_RF} can be used to represent the RF field in the linear optical and correlation analysis. We note that this effective photonic state has already successfully accounted for tunable high-order coherence in RF--laser interference~\cite{Yin2026OL} and Bell-inequality violations in two distinct RF schemes~\cite{Zeng2025Bell,AMZI2026}.

\subsection {Interferometric routes to entanglement}
\begin{figure*}[!thp]
\centering
\includegraphics[width= 0.9\textwidth]{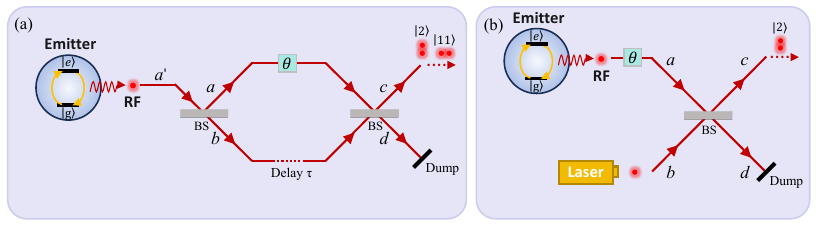} 
    \caption{Two interference schemes for synthesizing entangled states from RF.
    (a) Scheme A: self-interference of RF from a coherently driven two-level emitter in an AMZI. The input at port $a^\prime$ is split into a short arm $a$ with phase $\theta$ and a long arm $b$ with delay $\tau$, then recombined at the second BS, with output ports $c$ and $d$ and output state $\ket{\varphi}_{cd}$. The delay $\tau$ is chosen to be much longer than the emitter relaxation time, so that the two temporal modes can each be described by the steady-state emitter--field state of Eq.~\eqref{eq:joint_state}.
    (b) Scheme B: interference of RF with a coherent laser field $\ket{\mu}$ at a BS. The coherent amplitude and the relative phase \(\theta\) allow continuous control over the output state  \(|\Psi\rangle_{cd}\). Dashed arrows indicate subsequent routing to, e.g., a Franson interferometer for entanglement characterization.
    }
    \label{fig1}
\end{figure*}

Figure~\ref{fig1} illustrates two interferometric schemes for generating entangled states. Scheme A [Fig.~\ref{fig1}(a)], studied in Ref.~\cite{AMZI2026}, relies on self-homodyne of the RF field in an AMZI, where the two paths acquire a relative delay $\tau$ that is much longer than the emitter lifetime ($\tau \gg T_1$) and phase $\theta$ before recombination. Scheme B [Fig.~\ref{fig1}(b)], considered in the present work, instead interferes the RF field with a coherent laser state $|\mu\rangle$ at a beam splitter, with their relative phase $\theta$ and the laser amplitude providing independent control of the output state~\cite{LopezCarreno2018,Yin2026OL}.

We first consider Scheme A. As derived in Ref.~\cite{AMZI2026} using the computationally effective RF state given by Eq.~\eqref{eq:pure_RF}, the output state at ports $c$ and $d$ of the AMZI is
\begin{small}
\begin{equation}\label{eq1}
\begin{split}
|\varphi\rangle_{cd} = &\; p_0^{3/2}|00\rangle \\
& + \frac{e^{-i\theta}p_0\sqrt{p_1}}{\sqrt{2}} \Big[ (1+e^{i\theta})|01_c\rangle + (-1+e^{i\theta})|01_d\rangle + (1+e^{i\theta})|1_c0\rangle + (1-e^{i\theta})|1_d0\rangle \Big] \\
& + \frac{e^{-i\theta}\sqrt{p_0}p_1}{\sqrt{2}} \Big( |02_c\rangle - |02_d\rangle + |2_c0\rangle - |2_d0\rangle \Big) \\
& + \frac{e^{-2i\theta}\sqrt{p_0}p_1}{2} \Big[ (1+e^{i\theta}+e^{2i\theta})|1_c1_c\rangle + (1-e^{i\theta}+e^{2i\theta})|1_d1_d\rangle \\
& \quad + (-1-e^{i\theta}+e^{2i\theta})|1_c1_d\rangle + (-1+e^{i\theta}+e^{2i\theta})|1_d1_c\rangle \Big] \\
& + \frac{e^{-i\theta}p_1^{3/2}}{2} \Big( |1_c2_c\rangle - |1_c2_d\rangle + |1_d2_c\rangle - |1_d2_d\rangle \Big) \\
& + \frac{e^{-2i\theta}p_1^{3/2}}{2} \Big( -|2_c1_d\rangle + |2_c1_c\rangle + |2_d1_d\rangle - |2_d1_c\rangle \Big),
\end{split}
\end{equation}
\end{small}

\noindent where $|n_\alpha m_\beta\rangle$ denotes $n$ photons in the early temporal mode at port $\alpha$ and $m$ photons in the late temporal mode at port $\beta$, with $\alpha,\beta \in \{c,d\}$. This time-bin distinction is necessary because the AMZI introduces a relative delay $\tau$ between the long and short arms. For brevity, the common phase factor $e^{-i\omega_L t}$ has been omitted in the derivation. We write the vacuum as $0$ when the corresponding mode is empty. 
For example, $|2_c0\rangle$ means two photons in the early mode at port $c$ and vacuum in the late mode, while $|02_c\rangle$ means vacuum in the early mode and two photons in the late mode at port $c$.

With this notation, Eq.~\eqref{eq1} shows that at $\theta=\pi$ the single-photon terms
$|01_c\rangle$ and $|1_c0\rangle$ vanish at port $c$, since their amplitudes contain the
factor $1+e^{i\theta}$. The two-photon contributions then dominate, comprising both the
simultaneous pairs $|2_c0\rangle$ and $|0_c2_c\rangle$ and the time-separated pair
$|1_c1_c\rangle$. 
The entanglement in Scheme A is accessible through temporal post-selection, since the simultaneous $|2\rangle$ and time-separated $|11\rangle$ sectors coexist and are distinguished only by their coincidence-time difference~\cite{AMZI2026}. A cleaner source would retain the simultaneous two-photon component while eliminating the $|11\rangle$ sector, thereby producing entanglement from a single $|2\rangle$ sector. As shown below, this can be achieved by interfering RF with a coherent laser field.

\subsection{Interference between resonance fluorescence and a coherent field}

In Scheme~B, the RF state of Eq.~\eqref{eq:pure_RF} is homodyned with a coherent laser state $|\mu\rangle$ at a 50/50 beam splitter: the two fields share the same central frequency and are phase-locked with relative phase $\theta$.
 We define the relative field amplitude
$F=\sqrt{\mu p_0/p_1}$ and, for compactness, $\lambda=F\sqrt{p_1/(2p_0)}$. Using the general expression derived in Ref.~\cite{Yin2026OL}, the output state at ports $c$ and $d$ becomes
\begin{small}
\begin{align}\label{eq2}
\begin{split}
|\Psi\rangle_{cd} = &\; e^{-\frac{F^2 p_1}{2p_0}} \sqrt{p_0} \, |0\rangle \\
& + e^{-\frac{F^2 p_1}{2p_0}} \sqrt{\frac{p_1}{2}}
   \sum_{k=0}^{\infty} \frac{\lambda^k}{\sqrt{(k+1)!}}
   \Bigg[
      \left( F + e^{i\theta}(k+1) \right) |k+1\rangle_c |0\rangle_d
      + \left( F - e^{i\theta}(k+1) \right) |0\rangle_c |k+1\rangle_d
   \Bigg] \\
& + e^{-\frac{F^2 p_1}{2p_0}} \sqrt{\frac{p_1}{2}}
   \sum_{k=1}^{\infty} \sum_{s=0}^{k-1} 
   \frac{\lambda^k}{\sqrt{(s+1)! (k-s)!}}
   \left[ F + e^{i\theta}(2s+1-k) \right]
   |s+1\rangle_c |k-s\rangle_d ,
\end{split}
\end{align}
\end{small}

\noindent where $|s+1\rangle_c|k-s\rangle_d$ denotes a total of $n=k+1$ photons in the same temporal mode, with $s+1$ photons at port $c$ and $k-s$ photons at port $d$. 
The first summation describes all photons exiting through a single output port, while the second describes photons distributed between both ports, with at least one photon in each.

To expose the interference structure embedded in Eq.~\eqref{eq2}, we first consider the terms $|n\rangle_c|0\rangle_d$, in which all photons exit through port $c$. Their probability amplitudes are proportional to $A_n=F+ne^{i\theta}$, with $n\ge1$. At $\theta=\pi$, this reduces to $A_n=F-n$, giving a photon-number-selective cancellation condition at $F=n$.
At $F=1$, the coherent-field contribution to $A_1$ cancels that from the RF field, suppressing the single-photon component at port $c$. This is analogous to the first-order interference of fluorescence photons in Scheme~A: both schemes retain a two-photon contribution in the same temporal mode at a single output port.
As $F$ increases from 1 toward 2, the single-photon amplitude regains strength. At $F=2$, its coefficient is $A_1=2-1=1$, while the coherent-field and mixed RF--laser contributions to $A_2$ cancel, eliminating the $|2\rangle_c|0\rangle_d$ term. The resulting output can become strongly antibunched, with its nonvacuum population dominated by the single-photon component.

The cancellation conditions $A_n=0$ derived above are exact for the selected terms $|n\rangle_c|0\rangle_d$; the full photon-number distribution at port $c$ also receives contributions from the second summation, in which both output ports are populated. Nevertheless, these additional terms are negligible in the weak-drive limit ($p_0\to1$), so the above discussion captures the leading behaviour.

\begin{figure*}[!thp]
\centering
\includegraphics[width= 1.0\textwidth]{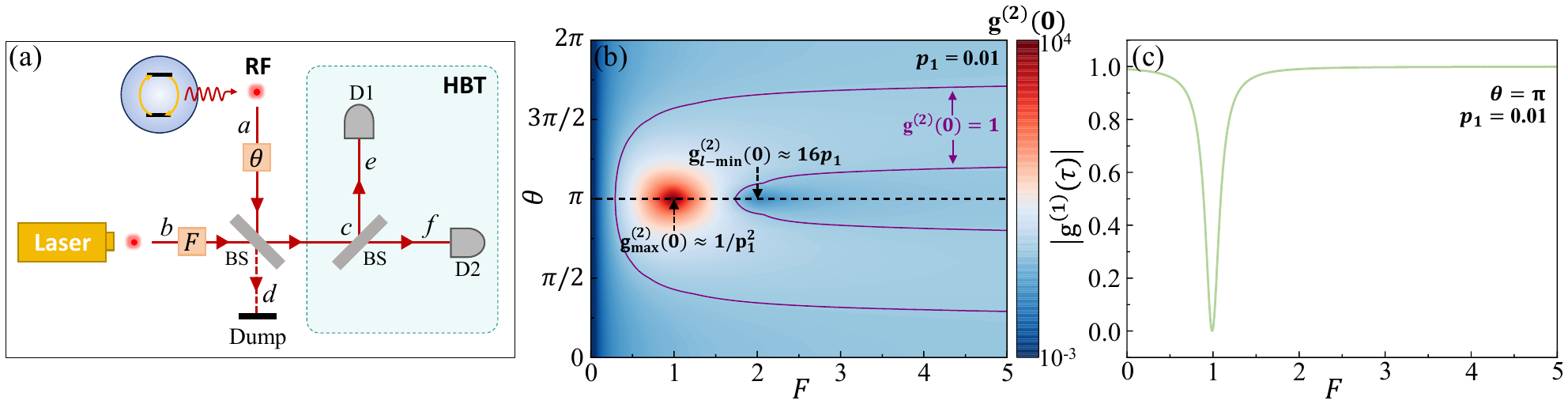}%
    \caption{Photon-number correlations and first-order coherence in RF--laser interference.
(a) Schematic of the second-order correlation measurement. The RF and coherent laser fields interfere at the first BS with relative phase \(\theta\) and amplitude \(F\). The interference output is directed to an HBT setup to measure \(g^{(2)}(0)\).
(b) Color map of the zero-delay second-order correlation function \(g^{(2)}(0)\) as a function of \(F\) and \(\theta\) for \(p_1=0.01\). The color scale is logarithmic. The purple curves mark the contour \(g^{(2)}(0)=1\), separating bunched and antibunched regions. The black dashed line indicates the \(\theta=\pi\) cross section, where the homodyne interference is strongest. Along this line, \(g^{(2)}(0)\) reaches a superbunching maximum near \((F,\theta)=(1,\pi)\), with \(g^{(2)}_{\max}(0)\approx1/p_1^2\), and a local antibunching minimum near \((F,\theta)=(2,\pi)\), with \(g^{(2)}_{l\text{-}\min}(0)\approx16p_1\).
(c) First-order coherence \(|g^{(1)}(\tau)|\) as a function of \(F\) for \(\theta=\pi\) and \(p_1=0.01\). 
}
    \label{fig2}
\end{figure*}

\subsection{Continuous control of photon-number statistics}

To examine how interference between the RF and coherent fields affects the photon statistics, we now evaluate the zero-delay second-order correlation function \(g^{(2)}(0)\) of the output field. As illustrated in Fig.~\ref{fig2}(a), the RF and the coherent laser are mixed at a BS, and the output channel $c$ is sent into a Hanbury Brown--Twiss (HBT) setup. The measurement records the single-count rate $C_1$ and the zero-delay coincidence rate $C_2$, from which \(g^{(2)}(0)=C_2/C_1^2\) is obtained, where $C_1^2$ is the baseline count rate for independent photons arrival.
To calculate \(g^{(2)}(0)\) theoretically, we write the joint quantum state of the laser and RF fields at the input ports:
\begin{equation}
|\Psi\rangle_{in} =|\varphi_a\rangle\otimes|\varphi_b\rangle = \left(\sqrt{p_0}|0\rangle_a + \sqrt{p_1}e^{i\theta}|1\rangle_a\right) \otimes |\mu\rangle_b,
\end{equation}
\noindent where $\sqrt{p_0}$ and $\sqrt{p_1}$ represent the magnitudes of the quantum probability amplitudes ($p_0+p_1=1$), \(|\mu\rangle = e^{-\mu/2}\sum_{n=0}^{\infty}(\sqrt{\mu})^n/\sqrt{n!}\,|n\rangle\) represents the coherent state with mean photon number \(\mu\), and \(\theta\) is the relative phase between the two input fields.

Firstly, we calculate the single-count rate $C_1$ and the zero-delay coincidence rate $C_2$ from the input states above (the detailed derivation is given in the Appendix A):
\begin{align}
C_1 &= \frac{p_1}{4} + \frac{\mu}{4} + \frac{\sqrt{p_0 p_1 \mu}}{2}\cos\theta, \label{eq4} \\
C_2 &= \frac{\mu^2}{16} + \frac{p_1\mu}{4} + \frac{\mu^{3/2}\sqrt{p_0 p_1}}{4}\cos\theta. \label{eq5}
\end{align}
Substituting Eqs.~(\ref{eq4}) and~(\ref{eq5}) into \(g^{(2)}(0)=C_2/C_1^2\) and using \(F = \sqrt{\mu p_0/p_1}\), we obtain
\begin{equation}
g^{(2)}(0)=
\frac{F^4 + 4F^2 p_0 + 4F^3 p_0 \cos\theta}
     {\left(p_0 + F^2 + 2F p_0 \cos\theta\right)^2}.
\label{eq6}
\end{equation}

Equation~\eqref{eq6} shows that the phase dependence of $g^{(2)}(0)$ arises entirely from terms containing $p_0$. Since $p_0$ also gives the long-delay first-order coherence of the RF field [Eq.~\eqref{eq:g1}], this phase sensitivity weakens as the driving power increases and $p_0$ decreases. In the strong-drive limit ($p_0\to0$), the effective RF state approaches a pure single-photon state and the phase dependence vanishes. This directly links the phase-dependent control of photon statistics to the first-order coherence of RF.

With Eq.~\eqref{eq6}, we analyze \(g^{(2)}(0)\) as a function of the amplitude \(F\) and phase \(\theta\) under a weak excitation of $p_1=0.01$. 
This choice retains a high degree of first-order coherence, $|g^{(1)}|=0.99$, which is essential for effective homodyne interference, while maintaining a practically useful photon flux. For example, for an emitter with $T_1=56~\mathrm{ps}$~\cite{Wu2023Optica}, $p_1=0.01$ corresponds to an emitted RF photon flux of approximately $9\times10^7~\mathrm{s^{-1}}$, or a collected flux of about $9\times10^6~\mathrm{s^{-1}}$ for a modest collection efficiency of $10\%$.

Figure~\ref{fig2}(b) presents \(g^{(2)}(0)\) as a two-dimensional color map over the relative field amplitude \(F\) and relative phase \(\theta\) for \(p_1=0.01\). The map is symmetric about \(\theta=\pi\), where destructive homodyne interference is strongest. The purple contours mark \(g^{(2)}(0)=1\), separating regions with bunched statistics, \(g^{(2)}(0)>1\), from those with antibunched statistics, \(g^{(2)}(0)<1\). At \(F=0\), corresponding to pure RF input, \(g^{(2)}(0)=0\) for all \(\theta\), reflecting the antibunched character of RF. As \(F\) increases from zero, the coherent-field contribution becomes more significant and the statistics generally evolve toward the Poissonian limit, \(g^{(2)}(0)=1\).

The most pronounced features occur along the \(\theta=\pi\) cross section, indicated by the black dashed line. Near \(F=1\), the coherent-field amplitude is matched to that of the RF field, so that the single-photon contributions cancel. This leaves the two-photon component relatively enhanced and produces a superbunching maximum, with \(g^{(2)}_{\max}(0)\approx1/p_1^2\). Near \(F=2\), the destructive-interference condition shifts to the two-photon component. Cancellation of this component produces a local antibunching minimum, with \(g^{(2)}_{l\text{-}\min}(0)\approx16p_1\). Although this antibunching resembles the pure-RF case at \(F=0\), its physical origin is different: here it results from interference-induced cancellation of the two-photon amplitude while the single-photon component remains finite.

In the weak-excitation limit \(p_0\to1\), the nonzero stationary points along \(\theta=\pi\) are exactly \(F=1\) and \(F=2\). For finite \(p_0\), they are shifted to
\[
F=\frac{3p_0\pm\sqrt{p_0(8-7p_0)}}{2(2p_0-1)}
\]
for \(p_0>1/2\). Thus, the superbunching maximum and antibunching minimum exist as distinct extrema only when \(p_0>1/2\); for \(p_0\leq1/2\), the antibunching minimum disappears. The dashed \(\theta=\pi\) line therefore identifies the most relevant operating condition for the following analysis.

It is useful to examine the first-order coherence, because residual phase coherence can contribute to the coincidence baseline in the subsequent Franson histogram~\cite{Zeng2025Bell}.
Figure~\ref{fig2}(c) shows \(|g^{(1)}(\tau)|\) as a function of \(F\) under weak excitation (\(p_1=0.01\)) and $\theta = \pi$, calculated using the AMZI scheme (see Appendix~B):
\begin{equation}
|g^{(1)}(\tau)|_{\theta=\pi}
=
\frac{(F-p_0)^2}
{(F-p_0)^2+p_0(1-p_0)}.
\end{equation}

At \(F=0\), corresponding to the absence of the laser field, this expression reduces to Eq.~\eqref{eq:g1}, with \(|g^{(1)}(\tau)|=p_0\). As \(F\) increases, the coherence exhibits a pronounced dip around \(F=1\), falling to exactly zero at \(F=p_0\), where the phase-coherent contributions from the RF and laser fields cancel. Beyond this point, the coherence recovers and approaches unity as the laser field increasingly dominates.

\begin{figure*}[!thp]
\centering
\includegraphics[width= 1.0\textwidth]{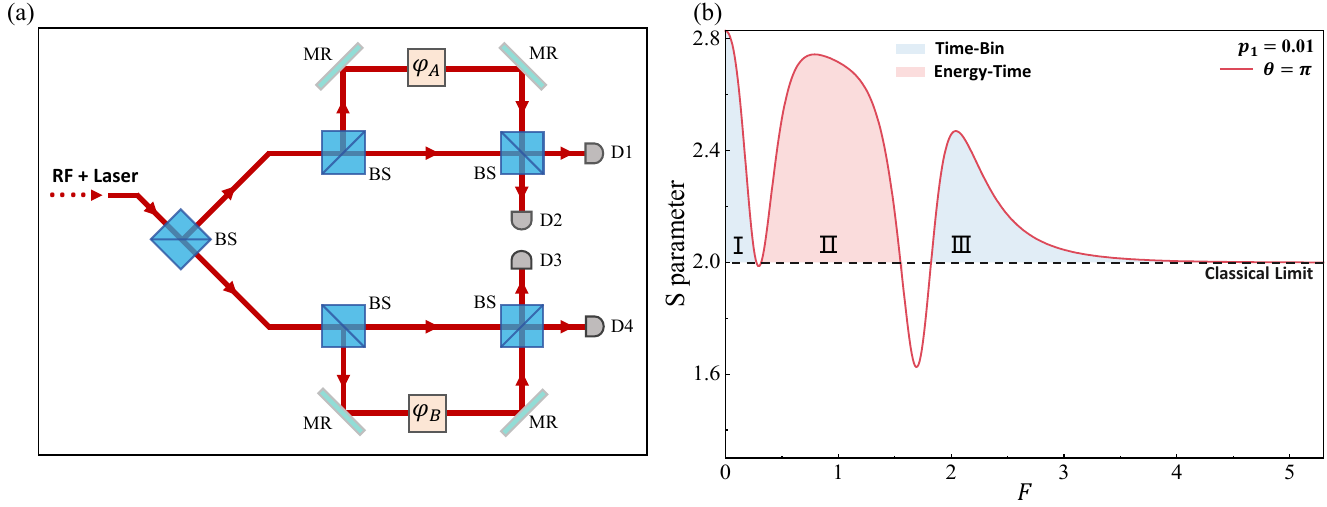}
    \caption{Analysis of the CHSH Bell parameter for RF--laser interference.
(a) Franson-interferometer configuration. The RF--laser interference field is directed into two analysis AMZIs with relative phases $\varphi_A$ and $\varphi_B$. All four output ports are monitored by single-photon detectors, allowing simultaneous measurement of the four coincidence combinations used to construct the CHSH parameter $S$.
(b) $S$ as a function of the relative field amplitude $F$ for $\theta=\pi$ and $p_1=0.01$, where the black dashed line indicates the classical limit $S=2$. The shaded regions I and III correspond to time-bin entanglement, while region II corresponds to energy--time entanglement. 
}        \label{fig3}
\end{figure*}

\subsection{Franson interference and Bell nonlocality}

Figure~\ref{fig3}(a) shows the Franson-interferometer configuration used to probe the entanglement of the RF--laser interference field. The field is divided between two analysis AMZIs, each introducing a temporal delay $\tau\gg T_1$ and a controllable phase $\varphi_A$ or $\varphi_B$, where $T_1$ is the radiative lifetime of the two-level emitter. The two output ports of each AMZI are monitored by single-photon detectors, allowing the four coincidence combinations required for the CHSH analysis to be recorded.

The zero-delay coincidence rate for a given pair of interferometer phases is (see Appendix~C)
\begin{small}
\begin{equation}\label{eq:C_Franson}
\begin{split}
\mathcal{C}(0)(F,\theta,\varphi_A,\varphi_B) = \frac{p_1^2}{256}\Bigg[
& \Big(2+2\cos(\varphi_A-\varphi_B)\Big) \\
& + 8F\cos\theta \Big(1+\cos\varphi_A+\cos\varphi_B+\cos(\varphi_A-\varphi_B)\Big) \\
& + F^2 \Bigg[ \left(\frac{12}{p_0}+4+4\cos 2\theta\right) 
      + \left(\frac{8}{p_0}+8+4\cos 2\theta\right)(\cos\varphi_A+\cos\varphi_B) \\
& \qquad\qquad + \left(\frac{4}{p_0}+4+4\cos 2\theta\right)\cos(\varphi_A-\varphi_B) 
      + 8\cos(\varphi_A+\varphi_B) \Bigg] \\
& + \frac{2F^3}{p_0} \Bigg[ 8\cos\theta \big(1+\cos\varphi_A+\cos\varphi_B\big) 
      + 4\cos\theta \Big(\cos(\varphi_A-\varphi_B)+\cos(\varphi_A+\varphi_B)\Big) \Bigg] \\
& + \frac{4F^4}{p_0^2}(1+\cos\varphi_A)(1+\cos\varphi_B)
\Bigg].
\end{split}
\end{equation}
\end{small}

From these coincidence rates, we define the correlation function~\cite{Clauser1969PRL}
\begin{equation}
E(\varphi_A,\varphi_B)=
\frac{
\mathcal{C}(0)(\varphi_A,\varphi_B)
+\mathcal{C}(0)(\varphi_A+\pi,\varphi_B+\pi)
-\mathcal{C}(0)(\varphi_A+\pi,\varphi_B)
-\mathcal{C}(0)(\varphi_A,\varphi_B+\pi)}
{
\mathcal{C}(0)(\varphi_A,\varphi_B)
+\mathcal{C}(0)(\varphi_A+\pi,\varphi_B+\pi)
+\mathcal{C}(0)(\varphi_A+\pi,\varphi_B)
+\mathcal{C}(0)(\varphi_A,\varphi_B+\pi)
}.
\end{equation}

The CHSH Bell parameter is then
\begin{equation}
S=
\left|
E(\varphi_A,\varphi_B)
-E(\varphi_A,\varphi_B')
+E(\varphi_A',\varphi_B)
+E(\varphi_A',\varphi_B')
\right|,
\end{equation}
where $(\varphi_A,\varphi_A')$ and $(\varphi_B,\varphi_B')$ are the two phase settings of the respective analysis interferometers (see Appendix~D). The CHSH inequality imposes the classical bound $S\leq2$, while $S>2$ signifies Bell-inequality violation.


To reveal entanglement crossovers, we first examine the simplest analytical limit \(p_0\to1\) at the maximally interfering phase \(\theta=\pi\). In this limit, the absolute coincidence rate is proportional to \(p_1^2\), but its normalized phase dependence, and hence the CHSH Bell parameter, remain finite and well defined. The essential evolution is captured by four representative values of the relative amplitude: \(F=0\), \(F=1\), \(F=2\), and \(F\to\infty\).

At \(F=0\), corresponding to pure RF, Eq.~\eqref{eq:C_Franson} reduces to
\begin{equation}
\left.\mathcal{C}(0)(\varphi_A,\varphi_B)\right|_{F=0}
=
\frac{p_1^2}{128}
\left[1+\cos(\varphi_A-\varphi_B)\right].
\end{equation}

The coincidence rate depends on the phase difference \(\varphi_A-\varphi_B\), arising from interference between the indistinguishable long--short and short--long paths. This corresponds to time-bin entanglement and yields the maximal Bell violation \(S=2\sqrt{2}\).

At \(F=1\), destructive interference cancels the single-photon component and the simultaneous two-photon component becomes dominant. The coincidence rate becomes
\begin{equation}
\left.\mathcal{C}(0)(\varphi_A,\varphi_B)\right|_{F=1}
=
\frac{p_1^2}{128}
\left[1+\cos(\varphi_A+\varphi_B)\right].
\end{equation}

The dependence now changes from the phase difference to the phase sum \(\varphi_A+\varphi_B\), corresponding to energy--time entanglement generated by interference between the short--short and long--long two-photon paths. The Bell parameter again reaches \(S=2\sqrt{2}\).

At \(F=2\), the simultaneous two-photon component is canceled and the field becomes strongly antibunched. The coincidence rate returns to
\begin{equation}
\left.\mathcal{C}(0)(\varphi_A,\varphi_B)\right|_{F=2}
=
\frac{p_1^2}{128}
\left[1+\cos(\varphi_A-\varphi_B)\right],
\end{equation}
recovering the same phase-difference dependence as for pure RF. The system therefore returns to time-bin entanglement, with \(S=2\sqrt{2}\).

Finally, in the limit \(F\to\infty\), the coherent field dominates and the coincidence rate factorizes as
\begin{equation}
\mathcal{C}(0)(\varphi_A,\varphi_B)|_{F\to\infty}
\propto
(1+\cos\varphi_A)(1+\cos\varphi_B).
\end{equation}

The responses of the two analysis interferometers are then independent, corresponding to classical coherent-state interference, and the Bell parameter approaches \(S=2\).

Together, these four limits reveal the sequence of regimes controlled by the relative coherent-field amplitude \(F\): time-bin entanglement at \(F=0\), energy--time entanglement at \(F=1\), recovered time-bin entanglement at \(F=2\), and finally classical coherent-field interference in the limit \(F\to\infty\). The first three regimes correspond to entangled two-photon correlations, while the last represents the loss of nonlocal correlations as the coherent field dominates.

Having established the limiting behavior, we now evaluate the full CHSH parameter for a practical weak excitation, \(p_1=0.01\) (\(p_0=0.99\)). Figure~\ref{fig3}(b) shows that the same sequence of regimes remains visible in the finite-excitation calculation, although the extrema of \(S\) are reduced relative to the ideal \(p_0\to1\) case, except at \(F=0\). Region~I corresponds to the RF-dominated time-bin regime. Starting from \(S=2\sqrt{2}\) at \(F=0\), the Bell parameter decreases rapidly and reaches a first valley near \(F\simeq0.30\), where it falls slightly below the classical bound \(S=2\). Region~II corresponds to the energy--time-dominated regime. In this region, \(S\) rises again and reaches a broad maximum of \(S\simeq2.74\) at \(F\simeq0.79\), close to the single-photon cancellation condition near \(F=1\). As \(F\) is increased further, \(S\) decreases to a second valley near \(F\simeq1.69\). Region~III corresponds to the recovered time-bin regime. Here, \(S\) increases again and reaches a smaller maximum of \(S\simeq2.47\) around \(F\simeq2.04\), close to the two-photon cancellation condition near \(F=2\). For still larger \(F\), the coherent field dominates and \(S\) gradually approaches the classical bound \(S=2\).

The displacement of the local maxima from exactly \(F=1\) and \(F=2\) does not change the underlying interference mechanism. These two values still identify the cancellation of the selected single- and two-photon amplitudes, respectively. At finite \(p_1\), however, the full Franson coincidence signal also contains additional Fock-state contributions from the laser input, which modify the optimized Bell parameter and shift the maxima of \(S\). Thus, Fig.~\ref{fig3}(b) should be viewed as a finite-excitation realization of the ideal crossover sequence discussed above, rather than as a different mechanism. A reference simulation in the ideal weak-excitation limit, $p_1=10^{-6}$, which more clearly reveals the crossover sequence and the corresponding Franson coincidence histograms, is provided in Appendix~E (Figure~A1).

The recovery of time-bin entanglement in Region~III is not unconditional. It relies on the antibunching associated with two-photon cancellation near \(F=2\). As discussed in Fig.~\ref{fig2}, this antibunching minimum exists only for \(p_0>1/2\). For \(p_0<1/2\), the antibunching minimum disappears, and the recovered time-bin-entangled regime is correspondingly lost.

\section{Conclusion}
In this work, we have revealed that the destructive interference between RF and a coherent laser field provides a viable mechanism for generating and continuously tuning entanglement. By adjusting the coherent amplitude \(F\), the system can be navigated across distinct entanglement regimes: from time-bin entanglement at \(F=0\), through energy-time entanglement 
around \(F=1\), back to time-bin entanglement 
around \(F=2\), and eventually to classical coherence as \(F\to\infty\). This crossover demonstrates that a single experimental configuration, based solely on passive linear optics and coherent field mixing, can realize qualitatively different classes of two-photon correlations through a single external parameter. The selective suppression of photon-number components via destructive interference is the underlying mechanism that enables this tunability, and it naturally extends beyond the two-photon level, suggesting a pathway toward higher-order multiphoton entangled states. Our results establish homodyned RF as a versatile and deterministic platform for controlled quantum state engineering, with potential implications for integrated photonic quantum information processing.

\section*{Acknowledgments}
{This research was funded by the Quantum Science and Technology-National Science and Technology Major Project (No. 2024ZD0302500),   
the National Natural Science Foundation of China under grants (12494604 and 12504410), and Beijing Academy of Quantum Information Sciences (Grant No. Y26NG262802).
}


\section*{Appendix A: Derivation of second-order correlation function  $g^{(2)}(0)$}
\renewcommand\theequation{A\arabic{equation}}
\setcounter{equation}{0}

As the set up shown in the Fig.~\ref{fig2}(a), the input state at time \(t\) is the product of two independent modes,
\begin{align}
    \ket{\Psi} &= \ket{\varphi_a}\otimes\ket{\varphi_b},\\
    \ket{\varphi_a} &= \sqrt{p_0}\,\ket{0}_a + \sqrt{p_1}\,e^{i\theta}\ket{1}_a,\qquad p_0+p_1=1,\\
    \ket{\varphi_b} &= e^{-\mu/2}\sum_{n=0}^{\infty}\frac{(\sqrt{\mu})^{n}}{\sqrt{n!}}\ket{n}_b,
\end{align}
where $\ket{\varphi_b}$ is a coherent state of real amplitude $\sqrt{\mu}$ (mean photon number $\mu = (p_1/p_0)\,F^{2}$, where the dimensionless parameter \(F\) controls the relative amplitude of the coherent field).
In the experiment, one output of a first 50:50 beam splitter is directed to a Hanbury Brown--Twiss (HBT) interferometer. 
The field operator of that port is $c = (a+b)/\sqrt{2}$, and we use the shorthand $a\equiv a_t$, $b\equiv b_t$. 
Inside the HBT setup, mode $c$ is further split by a second 50:50 beam splitter and the two outputs are monitored by single-photon detectors $D_1$ and $D_2$. 
For ideal detectors, the single-count rate of one detector is proportional to $\langle c^\dagger c\rangle/2$, while the coincidence rate between the two detectors is proportional to $\langle c^{\dagger 2}c^{2}\rangle/4$. 
It is therefore convenient to introduce the scaled operators
\begin{equation}
    \hat{O}_1 \equiv \frac{c}{\sqrt{2}} = \frac{a+b}{2},\qquad
    \hat{O}_2 \equiv \frac{c^{2}}{2} = \frac{a^{2}+2ab+b^{2}}{4},
\end{equation}
whose expectation values directly yield
\begin{equation}
    C_1 = \langle\Psi|\hat{O}_1^\dagger\hat{O}_1|\Psi\rangle = \frac12\langle c^\dagger c\rangle,\qquad
    C_2 = \langle\Psi|\hat{O}_2^\dagger\hat{O}_2|\Psi\rangle = \frac14\langle c^{\dagger 2}c^{2}\rangle .
\end{equation}

The normalized second-order correlation function at zero delay then takes the compact form
\begin{equation}
    g^{(2)}(0) = \frac{C_2}{C_1^{2}}.
\end{equation}

We first evaluate \(C_1\). For the single-photon operator \(\hat{O}_1=(a+b)/2\), using \(a\ket{\varphi_a}=\sqrt{p_1}e^{i\theta}\ket{0}_a\) and the coherent-state property \(b\ket{\varphi_b}=\sqrt{\mu}\ket{\varphi_b}\), we obtain
\begin{equation}
    \hat{O}_1\ket{\Psi} = \frac{1}{2}\Bigl[ \sqrt{p_1}e^{i\theta}\ket{0}_a\ket{\varphi_b} + \sqrt{\mu}\ket{\varphi_a}\ket{\varphi_b} \Bigr].
\end{equation}

Defining the normalized states \(\ket{X}=\ket{0}_a\ket{\varphi_b}\) and \(\ket{Y}=\ket{\varphi_a}\ket{\varphi_b}\), with \(\langle X|Y\rangle = \sqrt{p_0}\), we have
\begin{equation}
    \hat{O}_1\ket{\Psi} = \frac{\sqrt{p_1}}{2}e^{i\theta}\ket{X} + \frac{\sqrt{\mu}}{2}\ket{Y}.
\end{equation}

The single-count rate \(C_1 = |\hat{O}_1\ket{\Psi}|^{2}\) then becomes
\begin{align}
    C_1 &= \frac{p_1}{4}\langle X|X\rangle + \frac{\mu}{4}\langle Y|Y\rangle
          + 2\,\mathrm{Re}\Bigl[ \frac{\sqrt{p_1}}{2}e^{i\theta} \frac{\sqrt{\mu}}{2} \langle X|Y\rangle \Bigr] \\
        &= \frac{p_1}{4} + \frac{\mu}{4} + \frac{\sqrt{p_0 p_1 \mu}}{2}\cos\theta .
\end{align}

We now turn to \(C_2\). For the two-photon operator \(\hat{O}_2=(a^{2}+2ab+b^{2})/4\), since \(\ket{\varphi_a}\) contains at most one photon, \(a^{2}\ket{\varphi_a}=0\), so \(a^{2}\ket{\Psi}=0\). The cross terms give
\begin{equation}
    ab\ket{\Psi} = ba\ket{\Psi} = \sqrt{p_1\mu}\,e^{i\theta}\,\ket{0}_a\ket{\varphi_b} = \sqrt{p_1\mu}\,e^{i\theta}\ket{X},
\end{equation}
and the \(b^{2}\) term yields
\begin{equation}
    b^{2}\ket{\Psi} = \mu\,\ket{\varphi_a}\ket{\varphi_b} = \mu\ket{Y}.
\end{equation}

Assembling these results,
\begin{equation}
    \hat{O}_2\ket{\Psi} = \frac{\sqrt{p_1\mu}}{2}e^{i\theta}\ket{X} + \frac{\mu}{4}\ket{Y}.
\end{equation}

The coincidence rate \(C_2 = |\hat{O}_2\ket{\Psi}|^{2}\) then becomes
\begin{align}
    C_2 &= \frac{p_1\mu}{4}\langle X|X\rangle + \frac{\mu^{2}}{16}\langle Y|Y\rangle
          + 2\,\mathrm{Re}\Bigl[ \frac{\sqrt{p_1\mu}}{2}e^{i\theta} \frac{\mu}{4} \langle X|Y\rangle \Bigr] \\
        &= \frac{p_1\mu}{4} + \frac{\mu^{2}}{16} + \frac{\mu^{3/2}\sqrt{p_0 p_1}}{4}\cos\theta .
\end{align}

Substituting \(C_1\) and \(C_2\) into \(g^{(2)}(0)=C_2/C_1^{2}\) yields
\begin{equation}
     g^{(2)}(0) =
    \frac{ \dfrac{\mu^{2}}{16} + \dfrac{p_1\mu}{4} + \dfrac{\mu^{3/2}\sqrt{p_0 p_1}}{4}\cos\theta }
         { \left( \dfrac{p_1}{4} + \dfrac{\mu}{4} + \dfrac{\sqrt{p_0 p_1 \mu}}{2}\cos\theta \right)^{2} }.
\end{equation}

The interference terms in \(C_1\) and \(C_2\) scale as \(\sqrt{p_0 p_1\mu}\) and \(\mu^{3/2}\sqrt{p_0 p_1}\), respectively. In both cases, the presence of \(\cos\theta\) shows that the phase of the single-photon component in mode \(a\) can interfere with the coherent amplitude of mode \(b\), provided a vacuum component exists (\(p_0p_1>0\)). When mode \(a\) is a pure single-photon state (\(p_0=0\)), the interference vanishes and only the incoherent background remains. This behavior highlights the role of the vacuum amplitude in sustaining phase coherence in such hybrid quantum--classical states.

Substituting \(\mu = F^2 p_1/p_0\) into the above expression and multiplying both numerator and denominator by \(16 p_0^2 / p_1^2\), we obtain the simplified form
\begin{equation}
g^{(2)}(0)=
\frac{F^4 + 4F^2 p_0 + 4F^3 p_0 \cos\theta}
     {\left(p_0 + F^2 + 2F p_0 \cos\theta\right)^2},
\end{equation}
which is exactly Eq.~(\ref{eq6}) in the main text. For \(\theta=\pi\), the above equation reduces to
\begin{equation}\label{g2}
g^{(2)}(0)=\frac{F^4-4p_0F^3+4p_0F^2}{(F^2-2p_0F+p_0)^2}.
\end{equation}

Differentiating with respect to \(F\) and setting the derivative to zero gives
\begin{equation}
4p_0F\left[(2p_0-1)F^2-3p_0F+2p_0\right]=0.
\end{equation}

Apart from the trivial stationary point \(F=0\), which corresponds to the pure RF input with \(g^{(2)}(0)=0\), the non-zero stationary points satisfy
\begin{equation}
(2p_0-1)F^2-3p_0F+2p_0=0.
\end{equation}

Its solutions are
\begin{equation}
F=\frac{3p_0\pm\sqrt{p_0(8-7p_0)}}{2(2p_0-1)}
\qquad(p_0> 1/2).
\end{equation}

For \(p_0>1/2\), there are two positive stationary points. The smaller \(F\) corresponds to a maximum of \(g^{(2)}(0)\) (superbunching peak), while the larger \(F\) corresponds to a minimum (antibunching dip).
The extremal values of \(g^{(2)}(0)\) are obtained in closed form as
\begin{equation}
g^{(2)}_{\max}(p_0)
=
\frac{
2\left[
88p_0^3-163p_0^2+52p_0+24
-
\left(8p_0^2-41p_0+32\right)\Delta
\right]
}{
\left[
-4p_0^2-p_0+6
-
\left(5-4p_0\right)\Delta
\right]^2
},
\end{equation}
\begin{equation}
g^{(2)}_{\min}(p_0)
=
\frac{
2\left[
88p_0^3-163p_0^2+52p_0+24
+
\left(8p_0^2-41p_0+32\right)\Delta
\right]
}{
\left[
-4p_0^2-p_0+6
+
\left(5-4p_0\right)\Delta
\right]^2
},
\end{equation}
where
\begin{equation}
\Delta=\sqrt{p_0(8-7p_0)}.
\end{equation}

In the weak-excitation limit \(p_0\to 1\), writing \(p_1=1-p_0\ll 1\), the stationary-point equation reduces to \(F^2-3F+2=0\), i.e., \(F=1\) and \(F=2\), corresponding to the superbunching peak and antibunching dip, respectively, in agreement with Fig.~\ref{fig2}(b). Substituting these two values into above eqution, the extremal values of \(g^{(2)}(0)\) simplify to
\begin{equation}
g^{(2)}_{\max}\approx\frac{1}{p_1^2},\qquad
g^{(2)}_{\min}\approx16p_1.
\end{equation}

\section*{Appendix B: Calculation of \(|g^{(1)}(\tau)|\) versus \(F\) for $\theta=\pi$}
\renewcommand\theequation{B\arabic{equation}}
\setcounter{equation}{0}
Experimentally, first-order coherence function \(|g^{(1)}(\tau)|\) can be measured through an AMZI. We consider one AMZI of the Franson interferometer shown in Fig.~\ref{fig3}(a) of the main text, whose two arms have a relative time delay \(\tau\); its relative phase \(\phi_A\) is scanned, and the single counts \(C(\phi_A)\) at one output port are recorded. The normalized fringe contrast gives the interference visibility \(V\equiv |g^{(1)}(\tau)|\). The input state consists of the resonance-fluorescence field (mode \(a\)) and the coherent laser field (mode \(b\)), each containing early and late time bins, and is written as a direct product:
\begin{equation}
|\Psi_{\mathrm{in}}\rangle=|\phi_a\rangle_{t-\tau}\otimes
|\phi_a\rangle_t\otimes|\phi_b\rangle_{t-\tau}\otimes
|\phi_b\rangle_t,\label{eqstate}
\end{equation}
where
\begin{equation}
|\phi_a\rangle=\sqrt{p_0}|0\rangle+\sqrt{p_1}e^{i\theta}|1\rangle,\qquad
p_0+p_1=1,
\end{equation}
\begin{equation}
|\phi_b\rangle=e^{-\mu/2}
\sum_{n=0}^{\infty}
\frac{(\sqrt{\mu})^n}{\sqrt{n!}}
|n\rangle_b\equiv|\sqrt{\mu}\rangle .
\end{equation}

Here \(|\phi_a\rangle_{t-\tau}\) and \(|\phi_a\rangle_t\) denote the resonance-fluorescence field in the early and late time bins, respectively, while \(|\phi_b\rangle_{t-\tau}\) and \(|\phi_b\rangle_t\) denote the coherent laser field in the early and late time bins. Consider the operator
\begin{equation}
\hat{O}=a_{t-\tau}+b_{t-\tau}+e^{i\phi_A}a_t+e^{i\phi_A}b_t .
\end{equation}

The single-count rate is defined as
\begin{equation}
C(\phi_A)=\langle \Psi_{\mathrm{in}}|\hat{O}^\dagger \hat{O}|\Psi_{\mathrm{in}}\rangle=
|\hat{O}|\Psi_{\mathrm{in}}\rangle|^2 .
\end{equation}

For a term-by-term calculation, we write
\begin{equation}
\hat{O}=A+B+C+D,
\end{equation}
where
\begin{equation}
A=a_{t-\tau},\qquad
B=b_{t-\tau},\qquad
C=e^{i\phi_A}a_t,\qquad
D=e^{i\phi_A}b_t .
\end{equation}

Acting on the input state gives four terms:
\begin{equation}
T_1=A|\Psi_{\mathrm{in}}\rangle=
\sqrt{p_1}e^{i\theta}
|0\rangle_{a,t-\tau}
|\phi_a\rangle_t
|\phi_b\rangle_{t-\tau}
|\phi_b\rangle_t ,
\end{equation}
\begin{equation}
T_2=B|\Psi_{\mathrm{in}}\rangle=
\sqrt{\mu}
|\phi_a\rangle_{t-\tau}
|\phi_a\rangle_t
|\phi_b\rangle_{t-\tau}
|\phi_b\rangle_t ,
\end{equation}
\begin{equation}
T_3=C|\Psi_{\mathrm{in}}\rangle=
e^{i\phi_A}\sqrt{p_1}e^{i\theta}
|\phi_a\rangle_{t-\tau}
|0\rangle_{a,t}
|\phi_b\rangle_{t-\tau}
|\phi_b\rangle_t ,
\end{equation}
\begin{equation}
T_4=D|\Psi_{\mathrm{in}}\rangle=
e^{i\phi_A}\sqrt{\mu}
|\phi_a\rangle_{t-\tau}
|\phi_a\rangle_t
|\phi_b\rangle_{t-\tau}
|\phi_b\rangle_t .
\end{equation}

Therefore,
\begin{equation}
O|\Psi_{\mathrm{in}}\rangle=T_1+T_2+T_3+T_4 .
\end{equation}

The single-count rate is
\begin{equation}
C(\phi_A)=\sum_{i=1}^{4}\sum_{j=1}^{4}\langle T_i|T_j\rangle.
\end{equation}

Evaluating the 16 inner products term by term and using
\begin{equation}
\langle 0|\phi_a\rangle=\sqrt{p_0},\qquad
\langle \phi_a|\phi_a\rangle=1,\qquad
\langle \phi_b|\phi_b\rangle=1,
\end{equation}
we obtain
\begin{align}
\langle T_1|T_1\rangle &= p_1, &
\langle T_1|T_2\rangle &= \sqrt{p_0p_1\mu}\,e^{-i\theta}, \nonumber \\
\langle T_1|T_3\rangle &= p_0p_1 e^{i\phi_A}, &
\langle T_1|T_4\rangle &= \sqrt{p_0p_1\mu}\,e^{i(\phi_A-\theta)}, \nonumber \\
\langle T_2|T_1\rangle &= \sqrt{p_0p_1\mu}\,e^{i\theta}, &
\langle T_2|T_2\rangle &= \mu, \nonumber \\
\langle T_2|T_3\rangle &= \sqrt{p_0p_1\mu}\,e^{i(\phi_A+\theta)}, &
\langle T_2|T_4\rangle &= \mu e^{i\phi_A}, \nonumber \\
\langle T_3|T_1\rangle &= p_0p_1 e^{-i\phi_A}, &
\langle T_3|T_2\rangle &= \sqrt{p_0p_1\mu}\,e^{-i(\phi_A+\theta)}, \nonumber \\
\langle T_3|T_3\rangle &= p_1, &
\langle T_3|T_4\rangle &= \sqrt{p_0p_1\mu}\,e^{-i\theta}, \nonumber \\
\langle T_4|T_1\rangle &= \sqrt{p_0p_1\mu}\,e^{-i(\phi_A-\theta)}, &
\langle T_4|T_2\rangle &= \mu e^{-i\phi_A}, \nonumber \\
\langle T_4|T_3\rangle &= \sqrt{p_0p_1\mu}\,e^{i\theta}, &
\langle T_4|T_4\rangle &= \mu .
\end{align}

Summing these 16 terms and combining the conjugate pairs, we find
\begin{align}
C(\phi_A)
&=2p_1+2\mu+4\sqrt{p_0p_1\mu}\cos\theta \nonumber \\
&\quad+2(p_0p_1+\mu)\cos\phi_A+
4\sqrt{p_0p_1\mu}\cos\phi_A\cos\theta .
\end{align}

Using \(F=\sqrt{\mu p_0/p_1}\), we obtain
\begin{align}
C(\phi_A)&=2p_1\Bigg[1+\frac{F^2}{p_0}
+2F\cos\theta+
\left(p_0+\frac{F^2}{p_0}+2F\cos\theta\right)\cos\phi_A
\Bigg].
\end{align}

Writing
\begin{equation}
C(\phi_A)=A+B\cos\phi_A,
\end{equation}
we identify
\begin{equation}
A=2(1-p_0)\left[1+\frac{F^2}{p_0}+2F\cos\theta
\right],
\end{equation}
\begin{equation}
B=2(1-p_0)\left[p_0+\frac{F^2}{p_0}+2F\cos\theta\right].
\end{equation}

Since the scanned quantity is the phase \(\phi_A\), and the extracted visibility is
\begin{equation}
|g^{(1)}(\tau)|=V=\frac{C(F,0)-C(F,\pi)}{C(F,0)+C(F,\pi)}
=\frac{B}{A}.
\end{equation}

Substituting the expressions for \(A\) and \(B\) gives
\begin{equation}
|g^{(1)}(\tau)|=\frac{p_0+\dfrac{F^2}{p_0}+2F\cos\theta
}{1+\dfrac{F^2}{p_0}+2F\cos\theta}.
\end{equation}

Setting \(\theta=\pi\), so that \(\cos\theta=-1\), the above equation yields
\begin{equation}
|g^{(1)}(\tau)|_{\theta=\pi}=\frac{
p_0+\dfrac{F^2}{p_0}-2F}{1+\dfrac{F^2}{p_0}-2F}
=\frac{(F-p_0)^2}{(F-p_0)^2+p_0(1-p_0)},
\end{equation}
which is the general expression for \(|g^{(1)}(\tau)|\) as a function of \(F\) for \(\theta=\pi\). In particular, at \(F=p_0\), \(|g^{(1)}(\tau)|_{\theta=\pi}=0\), indicating that the first-order coherence vanishes completely. Using the above expression, we calculate the \(|g^{(1)}(\tau)|\) versus the relative field amplitude \(F\) for \(p_1=0.01\), as shown in Fig.~\ref{fig2}(c) of the main text. For \(F=0\),
\(
|g^{(1)}(\tau)|_{\theta=\pi}\big|_{F=0}=
\frac{p_0^2}{p_0^2+p_0(1-p_0)}=p_0 \).
For \(F=p_0\),
\(|g^{(1)}(\tau)|_{\theta=\pi}\big|_{F=p_0}=0.\)

\section*{Appendix C: Derivation of the coincidence 
rate}
\renewcommand\theequation{C\arabic{equation}}
\setcounter{equation}{0}
Considering the Franson interferometer shown in Fig.~\ref{fig3}(a) of the main text. The initial state consists of two spatial modes, each carrying two time bins labeled \(t-\tau\) (early) and \(t\) (late), which is the same as Eq.~(\ref{eqstate}).

The output measurement is described by the operator
\begin{equation}
    \hat{O} = \frac{1}{16}\bigl[\,\hat{a}_{t-\tau}+\hat{b}_{t-\tau}+e^{i\varphi_A}(\hat{a}_t+\hat{b}_t)\bigr]\,
                   \bigl[\,\hat{a}_{t-\tau}+\hat{b}_{t-\tau}+e^{i\varphi_B}(\hat{a}_t+\hat{b}_t)\bigr],
\end{equation}
which combines the two time bins with independently tunable phases \(\varphi_A,\varphi_B\); the prefactor \(1/16\) is a convenient normalization.

The action of individual operators on the initial state follows from the simple rules
\begin{equation}
    \hat{a}\ket{\varphi_a}=\sqrt{p_1}e^{i\theta}\ket{0},\quad \hat{a}^2\ket{\varphi_a}=0,\qquad
    \hat{b}\ket{\sqrt{\mu}}=\sqrt{\mu}\ket{\sqrt{\mu}},\quad \hat{b}^2\ket{\sqrt{\mu}}=\mu\ket{\sqrt{\mu}}
\end{equation}

Expanding \(\hat{O}\) generates sixteen terms, each being a product of two annihilation operators. The result of each operator product acting on \(\ket{\Psi_{\text{in}}}\) is summarized in Table~\ref{tab:sixteen}.  
(For brevity we use \(u=e^{i\varphi_A},\,v=e^{i\varphi_B}\).) We also introduce the normalized four-mode basis states to simplify the table
\begin{equation}
\begin{split}
    \ket{S_1} &= \ket{0}_{t-\tau}\ket{\varphi_a}_t\ket{\sqrt{\mu}}_{t-\tau}\ket{\sqrt{\mu}}_t, \\
    \ket{S_2} &= \ket{0}_{t-\tau}\ket{0}_t\ket{\sqrt{\mu}}_{t-\tau}\ket{\sqrt{\mu}}_t, \\
    \ket{S_3} &= \ket{\varphi_a}_{t-\tau}\ket{0}_t\ket{\sqrt{\mu}}_{t-\tau}\ket{\sqrt{\mu}}_t, \\
    \ket{S_4} &= \ket{\varphi_a}_{t-\tau}\ket{\varphi_a}_t\ket{\sqrt{\mu}}_{t-\tau}\ket{\sqrt{\mu}}_t.
\end{split}
\end{equation}

\begin{table}[h]
\centering
\caption{Result of each operator product acting on \(\ket{\Psi_{\text{in}}}\) (factor \(1/16\) omitted).}
\label{tab:sixteen}
\begin{tabular}{c|c}
Term & Result \\
\hline
1.\; \(\hat{a}_{t-\tau}\hat{a}_{t-\tau}\) & \(0\) \\
2.\; \(\hat{a}_{t-\tau}\hat{b}_{t-\tau}\) & \(\sqrt{\mu p_1}\,e^{i\theta}\,\ket{S_1}\) \\
3.\; \(\hat{a}_{t-\tau}(v\,\hat{a}_t)\)    & \(v\,p_1 e^{i2\theta}\,\ket{S_2}\) \\
4.\; \(\hat{a}_{t-\tau}(v\,\hat{b}_t)\)    & \(\sqrt{\mu p_1}\,e^{i\theta}\,v\,\ket{S_1}\) \\
5.\; \(\hat{b}_{t-\tau}\hat{a}_{t-\tau}\) & \(\sqrt{\mu p_1}\,e^{i\theta}\,\ket{S_1}\) \\
6.\; \(\hat{b}_{t-\tau}\hat{b}_{t-\tau}\) & \(\mu\,\ket{S_4}\) \\
7.\; \(\hat{b}_{t-\tau}(v\,\hat{a}_t)\)    & \(\sqrt{\mu p_1}\,e^{i\theta}\,v\,\ket{S_3}\) \\
8.\; \(\hat{b}_{t-\tau}(v\,\hat{b}_t)\)    & \(\mu\,v\,\ket{S_4}\) \\
9.\; \((u\,\hat{a}_t)\hat{a}_{t-\tau}\)    & \(u\,p_1 e^{i2\theta}\,\ket{S_2}\) \\
10.\; \((u\,\hat{a}_t)\hat{b}_{t-\tau}\)   & \(\sqrt{\mu p_1}\,e^{i\theta}\,u\,\ket{S_3}\) \\
11.\; \((u\,\hat{a}_t)(v\,\hat{a}_t)\)    & \(0\) \\
12.\; \((u\,\hat{a}_t)(v\,\hat{b}_t)\)    & \(\sqrt{\mu p_1}\,e^{i\theta}\,uv\,\ket{S_3}\) \\
13.\; \((u\,\hat{b}_t)\hat{a}_{t-\tau}\)   & \(\sqrt{\mu p_1}\,e^{i\theta}\,u\,\ket{S_1}\) \\
14.\; \((u\,\hat{b}_t)\hat{b}_{t-\tau}\)   & \(\mu\,u\,\ket{S_4}\) \\
15.\; \((u\,\hat{b}_t)(v\,\hat{a}_t)\)    & \(\sqrt{\mu p_1}\,e^{i\theta}\,uv\,\ket{S_3}\) \\
16.\; \((u\,\hat{b}_t)(v\,\hat{b}_t)\)    & \(\mu\,uv\,\ket{S_4}\)
\end{tabular}
\end{table}

Summing all contributions, the unnormalized output results takes the compact form
\begin{equation}
    \hat{O}\ket{\Psi_{\text{in}}} = \frac{1}{16}\bigl( C_1\ket{S_1}+C_2\ket{S_2}+C_3\ket{S_3}+C_4\ket{S_4} \bigr),
\end{equation}
with
\begin{align}
    C_1 &= \sqrt{\mu p_1}\,e^{i\theta}\,(2+u+v),\nonumber\\
    C_2 &= p_1 e^{i2\theta}\,(u+v),\nonumber\\
    C_3 &= \sqrt{\mu p_1}\,e^{i\theta}\,(u+v+2uv),\nonumber\\
    C_4 &= \mu\,(1+u)(1+v).
\end{align}

The experimentally measured coincidence count rate at the center zero delay is proportional to the norm squared of this unnormalized state. Denoting this rate by \(\mathcal{C}(0)\), we have
\begin{equation}
\begin{split}
    \mathcal{C}(0) &= |\hat{O}\ket{\Psi_{\text{in}}}|^2
    = \frac{1}{256}\Bigl[\,\sum_{i=1}^{4}|C_i|^2
        + 2\sqrt{p_0}\,\mathrm{Re}\bigl(C_1^*C_2+C_1^*C_4+C_2^*C_3+C_3^*C_4\bigr)
        \\&+ 2p_0\,\mathrm{Re}\bigl(C_1^*C_3+C_2^*C_4\bigr)\Bigr].
\end{split}        
\end{equation}

Here the inner products \(\langle S_i|S_j\rangle\) were evaluated using
\(\langle 0|\varphi_a\rangle=\sqrt{p_0}\) and \(\langle\varphi_a|\varphi_a\rangle=1\),
which yields the non‑trivial overlaps
\(\langle S_1|S_2\rangle=\langle S_1|S_4\rangle=\langle S_2|S_3\rangle=\langle S_3|S_4\rangle=\sqrt{p_0}\)
and \(\langle S_1|S_3\rangle=\langle S_2|S_4\rangle=p_0\).

Substituting the explicit coefficients \(C_1\dots C_4\) and expanding yields the fully trigonometric form

\begin{small}
\begin{align}
    \mathcal{C}(0)(\mu,\theta,\varphi_A,\varphi_B) = \frac{1}{256}\Big\{
&2\mu p_1\big(6 + 4\cos\varphi_A + 4\cos\varphi_B + 2\cos(\varphi_A-\varphi_B)\big) \nonumber\\
+&p_1^2\big(2 + 2\cos(\varphi_A-\varphi_B)\big) \nonumber\\
+&4\mu^2(1+\cos\varphi_A)(1+\cos\varphi_B) \nonumber\\
+&2p_0\mu p_1\big(2 + 4\cos\varphi_A + 4\cos\varphi_B + 2\cos(\varphi_A-\varphi_B) + 4\cos(\varphi_A+\varphi_B)\big) \nonumber\\
+&2\sqrt{p_0}\sqrt{\mu}\,p_1^{3/2}\,
\Big[ 2\cos(\theta+\varphi_A) + 2\cos(\theta+\varphi_B) + 2\cos\theta + 2\cos\theta\cos(\varphi_A-\varphi_B) \Big] \nonumber\\
+&2\sqrt{p_0}\sqrt{\mu}\,p_1^{3/2}\,
\Big[ 2(1+\cos(\varphi_A-\varphi_B))\cos\theta + 2\cos(\theta-\varphi_A) + 2\cos(\theta-\varphi_B) \Big] \nonumber\\
+&2\sqrt{p_0}\,\mu^{3/2}\sqrt{p_1}\,
\Big[ 4\cos\theta + 3\cos(\theta-\varphi_A) + 3\cos(\theta-\varphi_B) \nonumber\\
&\qquad\qquad\qquad + \cos(\theta+\varphi_A) + \cos(\theta+\varphi_B) + \cos(\theta-\varphi_A+\varphi_B) \nonumber\\
&\qquad\qquad\qquad + \cos(\theta+\varphi_A-\varphi_B) + 2\cos(\theta-\varphi_A-\varphi_B) \Big] \nonumber\\
+&2\sqrt{p_0}\,\mu^{3/2}\sqrt{p_1}\,
\Big[ 4\cos\theta + \cos(\theta-\varphi_A) + \cos(\theta-\varphi_B) \nonumber\\
&\qquad\qquad\qquad + 3\cos(\theta+\varphi_A) + 3\cos(\theta+\varphi_B) + \cos(\theta-\varphi_A+\varphi_B) \nonumber\\
&\qquad\qquad\qquad + \cos(\theta+\varphi_A-\varphi_B) + 2\cos(\theta+\varphi_A+\varphi_B) \Big] \nonumber\\
+&4p_0\mu p_1\big(1+\cos\varphi_A+\cos\varphi_B+\cos(\varphi_A-\varphi_B)\big)\cos 2\theta
\Big\}.
\label{eq:final}
\end{align}
\end{small}

Using the definition of mean photon number \(\mu = (p_1/p_0)\,F^{2}\), the above equation can be simplified as

\begin{small}
\begin{equation}
\begin{split}\label{Ccoin}
\mathcal{C}(0)(F,\theta,\varphi_A,\varphi_B) = \frac{p_1^2}{256}\Bigg[
& \Big(2+2\cos(\varphi_A-\varphi_B)\Big) \\
& + 8F\cos\theta \Big(1+\cos\varphi_A+\cos\varphi_B+\cos(\varphi_A-\varphi_B)\Big) \\
& + F^2 \Bigg[ \left(\frac{12}{p_0}+4+4\cos 2\theta\right) 
      + \left(\frac{8}{p_0}+8+4\cos 2\theta\right)(\cos\varphi_A+\cos\varphi_B) \\
& \qquad\qquad + \left(\frac{4}{p_0}+4+4\cos 2\theta\right)\cos(\varphi_A-\varphi_B) 
      + 8\cos(\varphi_A+\varphi_B) \Bigg] \\
& + \frac{2F^3}{p_0} \Bigg[ 8\cos\theta \big(1+\cos\varphi_A+\cos\varphi_B\big) 
      + 4\cos\theta \Big(\cos(\varphi_A-\varphi_B)+\cos(\varphi_A+\varphi_B)\Big) \Bigg] \\
& + \frac{4F^4}{p_0^2}(1+\cos\varphi_A)(1+\cos\varphi_B)
\Bigg].
\end{split}
\end{equation}
\end{small}

We first examine the case where the relative phase between the two input states is $\theta=\pi$. Setting $\cos\theta=-1$ and $\cos 2\theta=1$ yields

\begin{small}
\begin{equation}
\begin{split}
\mathcal{C}(0)(F,\varphi_A,\varphi_B) = \frac{p_1^2}{256}\Bigg[
& \big(2+2\cos(\varphi_A-\varphi_B)\big) \\
& -8F\big(1+\cos\varphi_A+\cos\varphi_B+\cos(\varphi_A-\varphi_B)\big) \\
& + F^2\Bigg[ \left(\frac{12}{p_0}+8\right) 
      + \left(\frac{8}{p_0}+12\right)(\cos\varphi_A+\cos\varphi_B) \\
&\qquad\qquad + \left(\frac{4}{p_0}+8\right)\cos(\varphi_A-\varphi_B) 
      + 8\cos(\varphi_A+\varphi_B) \Bigg] \\
& + \frac{2F^3}{p_0}\Bigg[ -8\big(1+\cos\varphi_A+\cos\varphi_B\big) 
      -4\big(\cos(\varphi_A-\varphi_B)+\cos(\varphi_A+\varphi_B)\big) \Bigg] \\
& + \frac{4F^4}{p_0^2}(1+\cos\varphi_A)(1+\cos\varphi_B)
\Bigg].
\end{split}
\end{equation}
\end{small}

With this phase convention fixed, we now examine the behavior of the coincidence rate as a function of the coherent amplitude \(F\). Specifically, we consider four representative limits: \(F=0\), corresponding to pure RF; \(F=1\), for which \(\mu=p_1/p_0\); \(F=2\), for which \(\mu=4p_1/p_0\); and the asymptotic limit \(F\to\infty\), where the coherent field dominates. These limits reveal the distinct physical regimes of the Franson interferometer.

1. For \(F=0\), only the zeroth-order term remains, the coincidence rate reduces to

\begin{equation}
\begin{split}
\mathcal{C}(0)(\varphi_A,\varphi_B) = \frac{p_1^2}{128} \big(1+\cos(\varphi_A-\varphi_B)\big)],
\end{split}
\end{equation}
which depends only on the phase difference \(\varphi_A-\varphi_B\). This result corresponds to the case where  the two RF photons are emitted at sequential times (early and late) and traverse the Franson interferometer via the short–long and long–short paths. The resulting \(\cos(\varphi_A-\varphi_B)\) dependence is characteristic of post-selected time-bin entanglement.

2. For \(F=1\), the coherent field has a mean photon number \(\mu = p_1/p_0\). The coincidence rate reduces to

\begin{equation}
\begin{split}
\mathcal{C}(0)(\varphi_A,\varphi_B) &= \frac{p_1^2}{256\,p_0^2}\Big[
(2p_0^2 -4p_0 +4) + (4p_0^2 -8p_0 +4)(\cos\varphi_A+\cos\varphi_B) \\&+ (2p_0^2 -4p_0 +2)\cos(\varphi_A-\varphi_B) + (8p_0^2 -8p_0 +2)\cos(\varphi_A+\varphi_B).
\end{split}
\end{equation}

In addition to the \(\cos(\varphi_A-\varphi_B)\) term from different-path interference, this expression also contains a \(\cos(\varphi_A+\varphi_B)\) term, which originates from two photons taking the same path (short–short and long–long). Their relative weights are modulated by the parameter \(p_0\).
When the vacuum component in the resonance-fluorescence field dominates (i.e., \(p_0\) is close to unity, corresponding to a weak excitation regime), the above equation simplifies dramatically to

\begin{equation}
\begin{split}
\mathcal{C}(0)(\varphi_A,\varphi_B)= \frac{p_1^2}{128}\big(1+\cos(\varphi_A+\varphi_B)\big),
\end{split}
\end{equation}
which shows that the interference pattern is governed solely by the sum phase \(\varphi_A+\varphi_B\), while the difference phase cancels out completely. This sum-phase dependence demonstrates two photon time-energy entanglement by a Franson interferometer.

3. Then, we consider the coincidence rate of the cases where the eigenvalue F = 2:

\begin{equation}
\begin{split}
\mathcal{C}(0)(\varphi_A,\varphi_B)  &= \frac{p_1^2}{256\,p_0^2}\Bigg[
(18p_0^2 - 80p_0 + 64)
+ (32p_0^2 - 96p_0 + 64)(\cos\varphi_A+\cos\varphi_B)\\&
+ (18p_0^2 - 48p_0 + 32)\cos(\varphi_A-\varphi_B)
+ (32p_0^2 - 64p_0 + 32)\cos(\varphi_A+\varphi_B)
\Bigg].
\end{split}
\end{equation}

In the weak-excitation limit (\(p_0 \to 1\)), the above equation reduces to
\[
\mathcal{C}(0)(\varphi_A,\varphi_B)  = \frac{p_1^2}{128} \left[1 + \cos(\varphi_A - \varphi_B)\right].
\]

This result shows that at $F=2$, the Franson interference pattern reverts to a phase-difference dependence, characteristic of time-bin entanglement and formally identical to the pure RF case at $F=0$. By contrast, at $F=1$ the phase-sum dependence signals energy-time entanglement. 

4. Finally, we consider the asymptotic limit of a very large coherent amplitude, \(F\to\infty\). In this regime, the dominant contribution in Eq.~(\ref{Ccoin}) comes from the \(F^4\) term

\begin{equation}
\begin{split}
\mathcal{C}(0)(\varphi_A,\varphi_B) \propto \frac{p_1^2 F^4}{64 p_0^2}(1+\cos\varphi_A)(1+\cos\varphi_B).
\end{split}
\end{equation}

This factorization indicates that in the strong laser drive, the two-photon coincidence rate becomes to the product of the two individual single-photon rate.

\section*{Appendix D: Analytical evaluation of the CHSH Bell parameter}
\renewcommand\theequation{D\arabic{equation}}
\setcounter{equation}{0}

Using the coincidence rate $\mathcal{C}(0)(F,\theta,\varphi_A,\varphi_B)$
derived in Eq.~(\ref{Ccoin}), we define the normalized correlation function as
\begin{small}
\begin{equation}
E(\varphi_A,\varphi_B)
=
\frac{
\mathcal{C}(0)(\varphi_A,\varphi_B)
+\mathcal{C}(0)(\varphi_A+\pi,\varphi_B+\pi)
-\mathcal{C}(0)(\varphi_A+\pi,\varphi_B)
-\mathcal{C}(0)(\varphi_A,\varphi_B+\pi)
}{
\mathcal{C}(0)(\varphi_A,\varphi_B)
+\mathcal{C}(0)(\varphi_A+\pi,\varphi_B+\pi)
+\mathcal{C}(0)(\varphi_A+\pi,\varphi_B)
+\mathcal{C}(0)(\varphi_A,\varphi_B+\pi)
}.
\label{eq:C1}
\end{equation}
\end{small}

The CHSH parameter is
\begin{equation}
S=
\left|
E(\varphi_A,\varphi_B)
-E(\varphi_A,\varphi'_B)
+E(\varphi'_A,\varphi_B)
+E(\varphi'_A,\varphi'_B)
\right|.
\label{eq:C2}
\end{equation}

Equation~(\ref{Ccoin}) can be rearranged as
\begin{equation}
\mathcal{C}(0)(\varphi_A,\varphi_B)
=
\frac{p_1^2}{256}
\left[
\mathcal{A}_0
+\mathcal{A}_1(\cos\varphi_A+\cos\varphi_B)
+\mathcal{A}_{-}\cos(\varphi_A-\varphi_B)
+\mathcal{A}_{+}\cos(\varphi_A+\varphi_B)
\right],
\label{eq:C3}
\end{equation}
where the coefficient $\mathcal{A}_1$ does not contribute to
Eq.~(\ref{eq:C1}), and
\begin{equation}
\begin{aligned}
\mathcal{A}_0={}&
2+8F\cos\theta
+F^2\left(
\frac{12}{p_0}+4+4\cos2\theta
\right)
+\frac{16F^3}{p_0}\cos\theta
+\frac{4F^4}{p_0^2},
\\
\mathcal{A}_{-}={}&
2+8F\cos\theta
+F^2\left(
\frac{4}{p_0}+4+4\cos2\theta
\right)
+\frac{8F^3}{p_0}\cos\theta
+\frac{2F^4}{p_0^2},
\\
\mathcal{A}_{+}={}&
8F^2
+\frac{8F^3}{p_0}\cos\theta
+\frac{2F^4}{p_0^2}.
\end{aligned}
\label{eq:C4}
\end{equation}

The local phase terms cancel in Eq.~(\ref{eq:C1}), giving
\begin{equation}
E(\varphi_A,\varphi_B)
=
\frac{
\mathcal{A}_{-}\cos(\varphi_A-\varphi_B)
+
\mathcal{A}_{+}\cos(\varphi_A+\varphi_B)
}{
\mathcal{A}_0
}.
\label{eq:C5}
\end{equation}

Writing Eq.~(\ref{eq:C5}) as
\begin{equation}
E(\varphi_A,\varphi_B)
=
T_x\cos\varphi_A\cos\varphi_B
+
T_y\sin\varphi_A\sin\varphi_B,
\end{equation}
with
\begin{equation}
T_x=
\frac{\mathcal{A}_{-}+\mathcal{A}_{+}}{\mathcal{A}_0},
\qquad
T_y=
\frac{\mathcal{A}_{-}-\mathcal{A}_{+}}{\mathcal{A}_0},
\end{equation}
the maximal CHSH value is
\begin{equation}
S_{\max}
=
2\sqrt{T_x^2+T_y^2}
=
\frac{
2\sqrt{2\left(
\mathcal{A}_{-}^2+\mathcal{A}_{+}^2
\right)}
}{
\mathcal{A}_0
}.
\label{eq:C6}
\end{equation}

Equation~(\ref{eq:C6}) therefore gives the maximal CHSH parameter
analytically for arbitrary $F$, $\theta$, and $p_0$ within the present model.

We now focus on the destructive-interference condition
$\theta=\pi$. In this case,
\begin{equation}
\mathcal{A}_{-}
=
\frac{2}{p_0^2}
\left(
F^2-2p_0F+p_0
\right)^2,
\qquad
\mathcal{A}_{+}
=
\frac{2F^2}{p_0^2}
\left(
F-2p_0
\right)^2,
\label{eq:C7}
\end{equation}
and
\begin{equation}
\mathcal{A}_0
=
\mathcal{A}_{-}
+\mathcal{A}_{+}
+\frac{8F^2(1-p_0)}{p_0}.
\label{eq:C8}
\end{equation}

Then the Eq.~(\ref{eq:C6}) reduces to
\begin{equation}
S_{\max}(F,p_0)
=
\frac{
2\sqrt{
2\left[
\left(F^2-2p_0F+p_0\right)^4
+
F^4\left(F-2p_0\right)^4
\right]
}
}{
\left(F^2-2p_0F+p_0\right)^2
+
F^2\left(F-2p_0\right)^2
+
4F^2p_0p_1
},
\end{equation}
where $p_0+p_1=1$. In the weak-excitation limit, $p_0\rightarrow1$, we define
\begin{equation}
w_{-}=(F-1)^4,
\qquad
w_{+}=F^2(F-2)^2.
\label{eq:C9}
\end{equation}

Equation~(\ref{eq:C5}) then reduces to
\begin{equation}
E(\varphi_A,\varphi_B)
=
\frac{
w_{-}\cos(\varphi_A-\varphi_B)
+
w_{+}\cos(\varphi_A+\varphi_B)
}{
w_{-}+w_{+}
}.
\label{eq:C10}
\end{equation}

Thus, $w_{-}$ and $w_{+}$ quantify the relative contributions of
the phase-difference and phase-sum correlations, respectively.
Equivalently,
\begin{equation}
E(\varphi_A,\varphi_B)
=
\cos\varphi_A\cos\varphi_B
+
\eta(F)\sin\varphi_A\sin\varphi_B,
\end{equation}
where
\begin{equation}
\eta(F)
=
\frac{
(F-1)^4-F^2(F-2)^2
}{
(F-1)^4+F^2(F-2)^2
}.
\label{eq:C11}
\end{equation}

The corresponding maximal Bell parameter is therefore
\begin{equation}
S_{\max}(F)
=
2\sqrt{1+\eta(F)^2}.
\label{eq:C12}
\end{equation}

For \(\theta=\pi\) and \(p_0\to1\), Eq.~(\ref{g2}) gives
\begin{equation}
g^{(2)}(0)=\frac{w_+}{w_-}.
\label{eq:D18}
\end{equation}

Thus \(\eta(F)\) can be expressed solely in terms of \(g^{(2)}(0)\):
\begin{equation}
\eta=\frac{1-g^{(2)}(0)}{1+g^{(2)}(0)}.
\label{eq:D19}
\end{equation}

Therefore, Eq.~(\ref{eq:C12}) reduces to
\begin{equation}
S_{\max}=2\sqrt{1+\left[\frac{1-g^{(2)}(0)}{1+g^{(2)}(0)}\right]^2
}.
\label{eq:D20}
\end{equation}

Equation~\eqref{eq:D20} establishes a direct connection between the CHSH parameter and the photon statistics in the weak-excitation limit.
When \(g^{(2)}(0)=0\), we have \(\eta=1\) and \(S_{\max}=2\sqrt{2}\). From \(g^{(2)}(0)=w_+/w_-=0\) we obtain \(F=0\) and \(F=2\) for \(p_0\to1\). The point \(F=0\) corresponds to pure RF input and is independent of \(p_1\).
When \(g^{(2)}(0)\to\infty\), i.e. \(w_-=0\), we have \(\eta\to-1\) and \(S_{\max}=2\sqrt{2}\). This occurs at \(F=1\) for \(p_0\to1\).
When \(g^{(2)}(0)=1\), we have \(\eta=0\) and \(S_{\max}=2\). From \(g^{(2)}(0)=1\), i.e. \(w_+=w_-\), we obtain \((F-1)^4=F^2(F-2)^2\), whose solutions are \(F_{\pm}=1\pm\frac{1}{\sqrt{2}}\simeq0.29,\;1.71\).
At these points, \(E(\phi_A,\phi_B)=\cos\phi_A\cos\phi_B\), so the correlation factorizes and \(S_{\max}=2\).
Thus, in the weak-excitation limit, the maxima of \(S\) occur at \(g^{(2)}(0)=0\) (\(F=0\) and \(F=2\)) and \(g^{(2)}(0)\to\infty\) (\(F=1\)), while the minima occur at \(g^{(2)}(0)=1\) (\(F\simeq0.29\) and \(1.71\)).

\section*{Appendix E: Reference simulation in the ideal weak-excitation limit}
\renewcommand\theequation{E\arabic{equation}}
\setcounter{equation}{0}

In the main text, we use \(p_1=0.01\) as a practical weak-excitation condition. This choice preserves high first-order coherence while maintaining an experimentally useful RF photon flux. To make the underlying interference mechanism more transparent, we also consider here a much weaker excitation condition with 
\begin{equation}
p_1=10^{-6},
\qquad
p_0\simeq1.    
\end{equation}

Using $p_1=\frac{2\Omega^2T_1^2}{1+2\Omega^2T_1^2}$,  this corresponds to
$\Omega T_1 \simeq 7.1\times10^{-4}$.
This regime is not intended as a practical operating point because of the low RF photon flux. Its purpose is instead to expose the limiting physics of RF--laser interference.

\begin{figure*}[!thp]
\centering
\includegraphics[width= 0.8\textwidth]{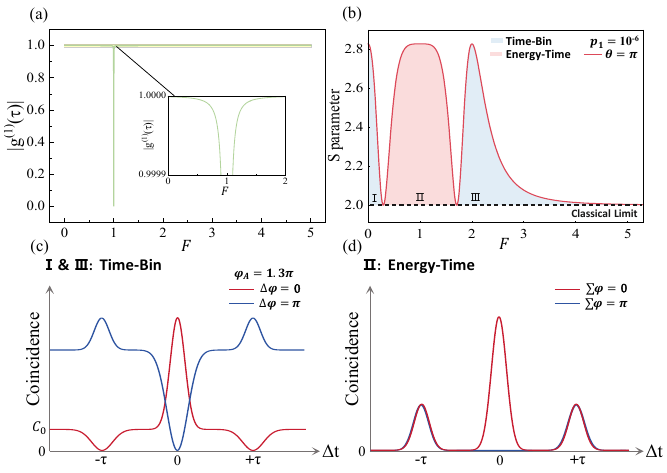}
\caption{Reference simulation for \(\theta=\pi\), with \(p_1=10^{-6}\). 
(a) \(|g^{(1)}(\tau)|\) as a function of  \(F\). Inset: a magnified view around $F =1$.
(b) CHSH Bell parameter \(S\) as a function of \(F\). 
(c) Representative Franson coincidence histogram for the time-bin entanglement regimes at \(F=0\) and \(F=2\), where the zero-delay rate  
depends on \(\Delta\varphi=\varphi_A - \varphi_B\).  The two curves correspond to \((\varphi_A,\varphi_B)=(1.3\pi,1.3\pi)\) and \((1.3\pi,0.3\pi)\), i.e., \(\Delta\varphi=0\) and \(\pi\). First-order coherence causes phase-dependent interference, so the baseline \(\mathcal{C}_0\) varies with the phase setting.
(d) Representative Franson coincidence histogram for the energy--time entanglement regime at \(F=1\), where the zero-delay coincidence rate 
depends on \(\Sigma\varphi=\varphi_A+\varphi_B\). The two curves correspond to \(\Sigma\varphi=0\) and \(\pi\). }
    \label{fig:appendix_weak}
\end{figure*}

Figure~\ref{fig:appendix_weak}(a) shows the first-order coherence \(|g^{(1)}(\tau)|\) for \(\theta=\pi\). Because \(p_0\simeq1\), the coherence remains essentially unity except in a very narrow region around \(F=p_0\simeq1\), where the phase-coherent RF and laser amplitudes cancel. This narrow dip illustrates the ideal homodyne cancellation condition more sharply than in the practical case \(p_1=0.01\).

Figure~\ref{fig:appendix_weak}(b) shows the CHSH Bell parameter \(S\) as a function of the relative field amplitude \(F\) for \(\theta=\pi\). In this ideal weak-excitation regime, the curve approaches the analytic limits discussed in the main text. The Bell parameter reaches values close to \(2\sqrt{2}\) at \(F=0\), \(F=1\), and \(F=2\), corresponding respectively to time-bin entanglement, energy--time entanglement, and recovered time-bin entanglement. Between these regimes, \(S\) falls to the classical bound near the points where \(g^{(2)}(0)=1\). For large \(F\), the coherent field dominates and \(S\) approaches the classical limit \(S=2\).

Figures~\ref{fig:appendix_weak}(c) and~\ref{fig:appendix_weak}(d) show representative Franson coincidence histograms expected for the two types of entanglement. For the time-bin entanglement regimes, corresponding to Regions~I and III, the phase-dependent two-photon interference is governed by the phase difference \(\Delta\varphi=\varphi_A-\varphi_B\). The relevant indistinguishable alternatives are the long--short and short--long paths of the two analysis AMZIs, giving coincidence profiles that switch between constructive and destructive interference as \(\Delta\varphi\) is varied. In these regimes, both the zero-delay coincidence probability and the baseline coincidence probability scale as \(p_1^2\). The baseline therefore remains finite relative to the zero-delay coincidence signal in the normalized histogram and, due to first-order interference, depends separately on the analyzer phases \(\varphi_A\) and \(\varphi_B\), rather than on their difference \(\Delta\varphi\)~\cite{Zeng2025Bell}.

By contrast, in the energy--time entanglement regime of Region~II, centered around \(F=1\), the single-photon component is canceled and the coincidence signal is dominated by the simultaneous two-photon component. The relevant indistinguishable alternatives are then the short--short and long--long paths, so that the interference depends on the phase sum \(\Sigma\varphi=\varphi_A+\varphi_B\). The zero-delay Franson coincidence probability scales as \(p_1^2\), whereas the baseline coincidence probability scales as \(p_1^4\). The baseline is therefore negligible in the weak-excitation limit. These histograms thus provide a direct time-domain distinction between phase-difference interference in the time-bin regimes and phase-sum interference in the energy--time regime. 

Overall, this idealized calculation provides a reference limit in which the three entangled regimes are clearly separated. It confirms that the finite-\(p_1\) results in the main text are a practical realization of the same underlying sequence: time-bin entanglement, energy--time entanglement, and recovered time-bin entanglement, followed by the approach to classical coherent-field interference.

\bibliography{reference}

@article{Brendel1999PRL,
  title = {Pulsed Energy-Time Entangled Twin-Photon Source for Quantum Communication},
  author = {Brendel, J. and Gisin, N. and Tittel, W. and Zbinden, H.},
  journal = {Phys. Rev. Lett.},
  volume = {82},
  issue = {12},
  pages = {2594--2597},
  numpages = {0},
  year = {1999},
  month = {Mar},
  publisher = {American Physical Society},
  doi = {10.1103/PhysRevLett.82.2594},
  url = {https://link.aps.org/doi/10.1103/PhysRevLett.82.2594}
}

@article{Bracht2026OpticaQ,
  author  = {Bracht, Thomas K. and Clark, Rachel N. and
             Androvitsaneas, Petros and Jordan, Matthew and
             Bishop, Samuel G. and Dyte, Harry E. and
             Cygorek, Moritz and Farrer, Ian A. and
             Reiter, Doris E. and Bennett, Anthony J.},
  title   = {Tunable multi-photon correlations from a coherently driven quantum dot},
  doi={10.1364/opticaq.580658},
  journal = {Opt. Quantum},
  volume  = {4},
  number  = {3},
  pages   = {224--231},
  year    = {2026}
}

@Article{Kim2025SA,
  author    = {Kim, Sang Kyu and Casalengua, Eduardo Zubizarreta and Boos, Katarina and Sbresny, Friedrich and Calcagno, Carolin and Riedl, Hubert and Finley, Jonathan J. and Antón-Solanas, Carlos and Laussy, Fabrice P. and Müller, Kai and Hanschke, Lukas and del Valle, Elena},
  journal   = {Sci. Adv.},
  title     = {Unlocking multiphoton emission from a single-photon source through mean-field engineering},
  doi = {10.1126/sciadv.adw3395},
  year      = {2025},
  number    = {44},
  pages     = {eadw3395},
  volume    = {11},
  comment   = {doi: 10.1126/sciadv.adw3395},
  publisher = {American Association for the Advancement of Science},
}

@misc{Zeng2025Bell,
      title={Bell Inequality Violation with Vacuum-One-Photon Number Superposition States}, 
      author={Zi-Qi Zeng and Jian Wang and Xiu-Bin Liu and Xu-Jie Wang and Li Liu and Hanqing Liu and Haiqiao Ni and Zhichuan Niu and Carlos Antón-Solanas and Bang Wu and Zhiliang Yuan},
      year={2025},
      eprint={2511.15413},
      archivePrefix={arXiv},
      primaryClass={quant-ph}
}

@article{Peiris2017PRL,
  author  = {Peiris, M. and Konthasinghe, K. and M\"uller, A.},
  title   = {Franson interference generated by a two-level system},
  journal = {Phys. Rev. Lett.},
  volume  = {118},
  number  = {3},
  pages   = {030501},
  year    = {2017},
  month = {Jan},
  publisher = {American Physical Society},
  doi = {10.1103/PhysRevLett.118.030501},
  url = {https://link.aps.org/doi/10.1103/PhysRevLett.118.030501}
}

@Article{LopezCarreno2018,
  author       = {López Carreño, J C and Casalengua, E Zubizarreta and Laussy, F P and del Valle, E},
  year         = {2018},
  journal = {Quantum Sci. Technol.},
  title        = {Joint subnatural-linewidth and single-photon emission from resonance fluorescence},
  doi = {10.1088/2058-9565/aacfbe},
  issn         = {2058-9565},
  number       = {4},
  pages        = {045001},
  volume       = {3},
  publisher    = {IOP Publishing},
}

@article{Zhong2018PRL,
  title = {12-Photon Entanglement and Scalable Scattershot Boson Sampling with Optimal Entangled-Photon Pairs from Parametric Down-Conversion},
  author = {Zhong, Han-Sen and Li, Yuan and Li, Wei and Peng, Li-Chao and Su, Zu-En and Hu, Yi and He, Yu-Ming and Ding, Xing and Zhang, Weijun and Li, Hao and Zhang, Lu and Wang, Zhen and You, Lixing and Wang, Xi-Lin and Jiang, Xiao and Li, Li and Chen, Yu-Ao and Liu, Nai-Le and Lu, Chao-Yang and Pan, Jian-Wei},
  journal = {Phys. Rev. Lett.},
  volume = {121},
  issue = {25},
  pages = {250505},
  numpages = {6},
  year = {2018},
  month = {Dec},
  publisher = {American Physical Society},
  month = {Dec},
  publisher = {American Physical Society},
  doi = {10.1103/PhysRevLett.121.250505},
  url = {https://link.aps.org/doi/10.1103/PhysRevLett.121.250505}
}

@Article{Knill2001Nature,
  author       = {Knill, E. and Laflamme, R. and Milburn, G. J.},
  year         = {2001},
  journal = {Nature},
  title        = {A scheme for efficient quantum computation with linear optics},
  doi          = {10.1038/35051009},
  issn         = {1476-4687},
  number       = {6816},
  pages        = {46--52},
  volume       = {409},
  refid        = {Knill2001},
}

@article{Bennett2000,
  title = {Quantum information and computation},
  author = {Bennett, Charles H. and DiVincenzo, David P.},
  journal = {Nature},
  volume = {404},
  number = {6775},
  pages = {247--255},
  year = {2000},
  doi = {10.1038/35005001},
}

@article{Clauser1969PRL,
  title = {{Proposed Experiment to Test Local Hidden-Variable Theories}},
  author = {Clauser, John F. and Horne, Michael A. and Shimony, Abner and Holt, Richard A.},
  journal = {Phys. Rev. Lett.},
  volume = {23},
  issue = {15},
  pages = {880--884},
  numpages = {0},
  year = {1969},
  month = {Oct},
  publisher = {American Physical Society},
  doi = {10.1103/PhysRevLett.23.880},
  url = {https://link.aps.org/doi/10.1103/PhysRevLett.23.880}
}

@article{Liu2024NP,
  title={Violation of {Bell} inequality by photon scattering on a two-level emitter},
  author={Liu, Shikai and Sandberg, Oliver August Dall’Alba and Chan, Ming Lai and Schrinski, Bj{\"o}rn and Anyfantaki, Yiouli and Nielsen, Rasmus B. and Larsen, Robert G. and Skalkin, Andrei and Wang, Ying and Midolo, Leonardo and Scholz, Sven and Wieck, Andreas D. and Ludwig, Arne and Sørensen, Anders S. and Tiranov, Alexey and Lodahl, Peter},
  doi          = {10.1038/s41567-024-02543-8},
  journal={Nat. Phys.},
  volume={20},
  number={9},
  pages={1429--1433},
  year={2024},
  publisher={Nature Publishing Group UK London}
}

@misc{AMZI2026,
      title={Engineering energy-time entanglement from resonance fluorescence}, 
      author={Jian Wang and Xiu-Bin Liu and Ziqi Zeng and Xu-Jie Wang and Carlos Antón-Solanas and Li Liu and Hanqing Liu and Haiqiao Ni and Zhichuan Niu and Bang Wu and Zhiliang Yuan},
      year={2026},
      eprint={2603.25341},
      archivePrefix={arXiv},
      primaryClass={quant-ph}
}

@article{Yin2026OL,
  title={Tunable high-order coherence in the interference of resonance fluorescence and laser light},
  doi = {10.1364/OL.592598},
  author={Yin, Hua-Lei and Liu, Rong-Zheng and Lyu, Sheng-Tao and Yuan, Zhiliang},
  journal={Opt. Let.},
  volume={51},
  number={9},
  pages={2616--2619},
  year={2026},
  publisher={Optica Publishing Group}
}

@article{Wang2025NC,
  author    = {Wang, Xu-Jie and Huang, Guoqi and Li, Ming-Yang and Wang, Yuan-Zhuo and Liu, Li and Wu, Bang and Liu, Hanqing and Ni, Haiqiao and Niu, Zhichuan and Ji, Weijie and Jiao, Rongzhen and Yin, Hua-Lei and Yuan, Zhiliang},
  title     = {Coherence in resonance fluorescence},
  doi      = {10.1038/s41467-025-61884-x},
  journal   = {Nat. Commun.},
  year      = {2025},
  volume    = {16},
  number    = {1},
  pages     = {6453},
  issn      = {2041-1723},
  date      = {2025-07-12},
}

@article{Bennett1993Teleportation,
  author  = {Bennett, Charles H. and Brassard, Gilles and Cr{\'e}peau, Claude and Jozsa, Richard and Peres, Asher and Wootters, William K.},
  title   = {Teleporting an Unknown Quantum State via Dual Classical and Einstein--Podolsky--Rosen Channels},
  journal = {Phys. Rev. Lett.},
  volume  = {70},
  pages   = {1895--1899},
  year    = {1993},
  month = {Mar},
  publisher = {American Physical Society},
  doi = {10.1103/PhysRevLett.70.1895},
  url = {https://link.aps.org/doi/10.1103/PhysRevLett.70.1895}
}

@article{Bennett1996Purification,
  author  = {Bennett, Charles H. and Brassard, Gilles and Popescu, Sandu and Schumacher, Benjamin and Smolin, John A. and Wootters, William K.},
  title   = {Purification of Noisy Entanglement and Faithful Teleportation via Noisy Channels},
  journal = {Phys. Rev. Lett.},
  volume  = {76},
  pages   = {722--725},
  year    = {1996},
  month = {Jan},
  publisher = {American Physical Society},
  doi = {10.1103/PhysRevLett.76.722},
  url = {https://link.aps.org/doi/10.1103/PhysRevLett.76.722}
}

@article{Ekert1991PRL,
  author  = {Ekert, Artur K.},
  title   = {Quantum Cryptography Based on Bell's Theorem},
  journal = {Phys. Rev. Lett.},
  volume  = {67},
  number  = {6},
  pages   = {661--663},
  year    = {1991},
  month = {Aug},
  publisher = {American Physical Society},
  doi = {10.1103/PhysRevLett.67.661},
  url = {https://link.aps.org/doi/10.1103/PhysRevLett.67.661}
}

@article{Duan2001Nature,
  author   = {Duan, L.-M. and Lukin, M. D. and Cirac, J. I. and Zoller, P.},
  title    = {Long-distance quantum communication with atomic ensembles and linear optics},
  doi      = {10.1038/35106500},
  issn     = {1476-4687},
  number   = {6862},
  pages    = {413--418},
  url      = {https://doi.org/10.1038/35106500},
  volume   = {414},
  journal  = {Nature},
  refid    = {Duan2001},
  year     = {2001},
}

@article{Kok2007RMP,
  author  = {Kok, Pieter and Munro, W. J. and Nemoto, Kae
             and Ralph, T. C. and Dowling, Jonathan P.
             and Milburn, G. J.},
  title   = {Linear optical quantum computing with photonic qubits},
  journal = {Rev. Mod. Phys.},
  volume  = {79},
  pages   = {135--174},
  year    = {2007},
  month = {Jan},
  publisher = {American Physical Society},
  doi = {10.1103/RevModPhys.79.135},
  url = {https://link.aps.org/doi/10.1103/RevModPhys.79.135}
}

@article{Giovannetti2011NP,
  author  = {Giovannetti, Vittorio and Lloyd, Seth and Maccone, Lorenzo},
  title   = {Advances in quantum metrology},
  doi          = {10.1038/nphoton.2011.35},
  journal = {Nature Photon.},
  volume  = {5},
  pages   = {222--229},
  year    = {2011}
}

@article{Kwiat1995PRL,
  author  = {Kwiat, Paul G. and Mattle, Klaus and Weinfurter, Harald
             and Zeilinger, Anton and Sergienko, Alexander V.
             and Shih, Yanhua},
  title   = {New High-Intensity Source of Polarization-Entangled Photon Pairs},
  journal = {Phys. Rev. Lett.},
  volume  = {75},
  pages   = {4337--4341},
  year    = {1995},
  month = {Dec},
  publisher = {American Physical Society},
  doi = {10.1103/PhysRevLett.75.4337},
  url = {https://link.aps.org/doi/10.1103/PhysRevLett.75.4337}
}

@article{Pan2012RMP,
  author  = {Pan, Jian-Wei and Chen, Zeng-Bing and Lu, Chao-Yang
             and Weinfurter, Harald and Zeilinger, Anton
             and {\.Z}ukowski, Marek},
  title   = {Multiphoton entanglement and interferometry},
  journal = {Rev. Mod. Phys.},
  volume  = {84},
  pages   = {777--838},
  year    = {2012},
  month = {May},
  publisher = {American Physical Society},
  doi = {10.1103/RevModPhys.84.777},
  url = {https://link.aps.org/doi/10.1103/RevModPhys.84.777}
}

@article{Huang2026PRA,
  title = {Two-photon interference between mutually detuned resonance-fluorescence signals scattered off a semiconductor quantum dot},
  author = {Huang, Guoqi and Wang, Jian and Zeng, Ziqi and Liu, Hanqing and Liu, Li and Ji, Weijie and Wu, Bang and Ni, Haiqiao and Niu, Zhichuan and Jiao, Rongzhen and Marangon, Davide G. and Yuan, Zhiliang},
  journal = {Phys. Rev. A},
  volume = {113},
  issue = {2},
  pages = {023718},
  numpages = {12},
  year = {2026},
  month = {Feb},
  publisher = {American Physical Society},
  doi = {10.1103/b6l5-8fdq},
  url = {https://link.aps.org/doi/10.1103/b6l5-8fdq}
}

@article{nguyen2011ultra,
	title={Ultra-coherent single photon source},
	author={Nguyen, Hai-Son and Sallen, Gregory and Voisin, Christophe and Roussignol, Ph and Diederichs, Carole and Cassabois, Guillaume},
	journal={Appl. Phys. Lett.},
	volume={99},
	number={26},
	pages={261904},
	year={2011},
	publisher={American Institute of Physics},
    doi      = {10.1063/1.3672034}
}

@article{mattiesen2012,
	title = {Subnatural Linewidth Single Photons from a Quantum Dot},
	author = {Matthiesen, Clemens and Vamivakas, Anthony Nickolas and Atat\"ure, Mete},
	journal = {Phys. Rev. Lett.},
	volume = {108},
	issue = {9},
	pages = {093602},
	numpages = {4},
	year = {2012},
  month = {Feb},
  publisher = {American Physical Society},
  doi = {10.1103/PhysRevLett.108.093602},
  url = {https://link.aps.org/doi/10.1103/PhysRevLett.108.093602}
}

@article{hoffges1997heterodyne,
	title={Heterodyne measurement of the fluorescent radiation of a single trapped ion},
	author={H{\"o}ffges, JT and Baldauf, HW and Eichler, T and Helmfrid, SR and Walther, H},
	journal={Opt. Commun.},
	volume={133},
	number={1-6},
	pages={170--174},
	year={1997},
	publisher={Elsevier},
    doi = {https://doi.org/10.1016/S0030-4018(96)00621-9}
}

@article{Franson1989,
	author = {J. D. Franson},
	title = {Bell Inequality for Position and Time},
	journal = {Phys. Rev. Lett.},
	volume = {62},
	pages = {2205--2208},
	year = {1989},
  month = {May},
  publisher = {American Physical Society},
  doi = {10.1103/PhysRevLett.62.2205},
  url = {https://link.aps.org/doi/10.1103/PhysRevLett.62.2205}
}

@article{Li2005,
  author  = {Li, Xiaoying and Voss, Paul L. and Sharping, Jay E. and Kumar, Prem},
  title   = {Optical-Fiber Source of Polarization-Entangled Photons in the 1550 nm Telecom Band},
  journal = {Phys. Rev. Lett.},
  volume  = {94},
  pages   = {053601},
  year    = {2005},
  month = {Feb},
  publisher = {American Physical Society},
  doi = {10.1103/PhysRevLett.94.053601},
  url = {https://link.aps.org/doi/10.1103/PhysRevLett.94.053601}
}

@article{Takesue2005,
  author  = {Takesue, Hiroki and Inoue, Kyo},
  title   = {Generation of 1.5-$\mu$m band time-bin entanglement using spontaneous fiber four-wave mixing and planar lightwave circuit interferometers},
  journal = {Phys. Rev. A},
  volume  = {72},
  pages   = {041804(R)},
  year    = {2005},
  month = {Oct},
  publisher = {American Physical Society},
  doi = {10.1103/PhysRevA.72.041804},
  url = {https://link.aps.org/doi/10.1103/PhysRevA.72.041804}
}

@article{Bouwmeester1997Nature,
  author  = {Bouwmeester, Dik and Pan, Jian-Wei and Mattle, Klaus
             and Eibl, Manfred and Weinfurter, Harald and Zeilinger, Anton},
  title   = {Experimental quantum teleportation},
  doi          = {10.1038/37539},
  journal = {Nature},
  volume  = {390},
  pages   = {575--579},
  year    = {1997}
}

@article{Walther2005Nature,
  author  = {Walther, Philip and Resch, Kevin J. and Rudolph, Terry
             and Schenck, Emmanuel and Weinfurter, Harald
             and Vedral, Vlatko and Aspelmeyer, Markus and Zeilinger, Anton},
  title   = {Experimental one-way quantum computing},
  doi          = {10.1038/nature03347},
  journal = {Nature},
  volume  = {434},
  pages   = {169--176},
  year    = {2005}
}

@article{Wu2023Optica,
  author  = {Wu, Bang and Wang, Xu-Jie and Liu, Li and Huang, Guoqi and Wang, Wenyan and Liu, Hanqing and Ni, Haiqiao and Niu, Zhichuan and Yuan, Zhiliang},
  title   = {Mollow Triplets under Few-Photon Excitation},
  doi = {10.1364/OPTICA.491565},
  journal = {Optica},
  volume  = {10},
  number  = {8},
  pages   = {1118--1123},
  year    = {2023}
}

\end{document}